# Phenomenology of the dark matter sector in the 2HDM extended with complex scalar singlet

**Juhi Dutta**[1,a], **Gudrid Moortgat-Pick**[1,2,b], **Merle Schreiber**[1,2,c]

[1] II. Institut für Theoretische Physik, Universität Hamburg, Luruper Chaussee 149, 22761 Hamburg, Germany
[2] Deutsches Elektronen-Synchrotron DESY, Notkestr. 85, 22607 Hamburg, Germany

Received: 4 September 2024 / Accepted: 13 January 2025

**Abstract** The two-Higgs-doublet model augmented with a complex scalar singlet (2HDMS) is a well-motivated candidate for Beyond Standard Model (BSM) Physics. We investigate the dark matter phenomenology of the 2HDMS with the complex scalar singlet as the dark matter candidate. We perform a study of the parameter space allowed by existing theoretical and experimental constraints from dark matter, flavour physics and collider searches. Further, we discuss a few benchmark scenarios to test the discovery potential for the 2HDMS at the HL-LHC and at future high-energy $e^+e^-$ colliders.

## 1 Introduction

The presence of dark matter (DM) has been unequivocally established from experimental observations via its gravitational interactions. However, it still remains a puzzle at the interface between particle physics and cosmology. In particular, the nature of dark matter has been under contemplation for several decades and particle physics models providing a possible dark matter candidate are being actively explored at both theoretical and experimental frontiers.

One possible particulate candidate for dark matter is a scalar singlet under the Standard Model (SM) gauge group added to the SM [1–4] and interacting with the SM particles via the 125 GeV Higgs boson as the portal to the dark sector. Such a minimal Higgs portal dark matter model is stringently constrained from DM direct detection experiments [5] such as LUX-ZEPLIN (LZ) [6], thus providing a strong motivation to look for non-minimal Higgs sectors with additional portals to the dark sector.

On the other hand, with the discovery of a 125 GeV SM-like Higgs scalar, the question arises whether the Higgs is the only scalar particle or part of a non-minimal scalar sector as is the case for fermions or gauge bosons. The simplest non-minimal Higgs sector arises in the two-Higgs-doublet model (2HDM). The 2HDM is a minimal extension of the SM by an extra SU(2) Higgs doublet. After electro-weak symmetry breaking (EWSB), there are five physical Higgs bosons, namely, two CP-even Higgses $h$, $H$, a pseudoscalar $A$ and a pair of charged Higgses $H^{\pm}$. Besides a non-minimal Higgs sector, it also provides the possibility of accommodating dark matter for special cases such as the inert doublet model (IDM) where the second doublet is inert under the SM gauge group. For a general 2HDM scenario, however, one requires extra particles to accommodate dark matter. A minimal extension of 2HDM with an extra scalar singlet provides a natural DM candidate. While real scalar singlet extensions of the 2HDM, have been well studied to address dark matter with the heavy CP-even Higgs bosons as portals to DM [7–13] and in some cases with discovery potential in the upcoming HL-LHC run [14, 15], complex scalar extensions to the 2HDM have been studied in the context of modified Higgs sectors [16–20], pseudo-Nambu Goldstone dark matter together with a mixed Higgs sector [17, 21], singlet-doublet and fermionic dark matter [5] as well as in connection with gravitational waves signals, and in the context of axion models such as the Dine-Fischler-Srednicki-Zhitnitsky (DFSZ) model [22, 23]. A continued search of singlet extended multi-Higgs models is warranted at future colliders, motivated by dark matter, neutrino masses, inflation, baryogenesis and also the possibility of accommodating current scalar excesses at colliders. While our study mainly focusses on dark matter phenomenology in 2HDMS, scalar singlet extensions have also been studied in the context of addressing other issues of nature such as baryogenesis (see [24] for the 2HDM extended with a real singlet scalar), neutrino masses and inflation [25]. Singlet extended 2HDMs have also been studied to explain some observed excesses at colliders such as the 95 GeV light scalar excess [17, 18, 20] at LEP and LHC as well as the neutrino anomalies [26], the (g-2) excess [27, 28] and the gamma-ray excess [11, 29].

While our work focuses on the complex scalar extension of the 2HDM in the context of dark matter, where the scalar singlet does not obtain a vacuum expectation value(*vev*) and contributes to a two-component dark matter scenario, we discuss some existing studies in comparison with our current model to highlight the differences with respect to this study. Reference [17] deals with S2HDM with a U(1) symmetry which is broken when the scalar part of the complex singlet scalar obtains a *vev* while the pseudo-Nambu Goldstone (pNG) boson contributing as a one-component DM candidate. In this case, since the singlet scalar does not obtain a *vev* and both scalar and pseudoscalar components are stable, it constitutes a two-component DM scenario. Further, Ref. [17] focusses on the pNG DM scenario with dark matter masses around the SM Higgs funnel region (with dark matter masses less than 100 GeV) and explored the effects of indirect detection for constraining the DM scenario along with a 95 GeV scalar excess. We have studied the case where the complex scalar mass is not limited to less than 100 GeV but also heavier dark matter with masses up to $\sim 900-1000$ GeV covering the WIMP region. In both models, the resonance features of the Higgses occur, i.e., we observe similarities in the Higgs 125 resonance region around DM mass 62–63 GeV. While Ref. [17] allows for the presence of an extra light Higgs which may explain the 95 GeV excess, in the current study, it is not possible for that in the 2HDMS two-component DM scenario for the Type II scenario due to the absence of an extra Higgs as the singlet scalar does not obtain a *vev*. Further, Ref. [20] has studied the case for 95 GeV excess in 2HDMS along with a pseudoscalar dark matter.

[a] e-mail: juhi.dutta@desy.de (corresponding author)
[b] e-mail: gudrid.moortgat-pick@desy.de
[c] e-mail: merle.schreiber@desy.de

In this work, we investigate a next-to-minimal extension of the Type II 2HDM with a complex scalar singlet (2HDMS) in the context of dark matter, where the complex scalar does not develop a *vev*[1]. This ensures that the Higgs doublet does not mix with the complex scalar resulting in the same Higgs sector as in the 2HDM, i.e., consisting of two CP-even Higgs bosons, $h, H$, a pseudoscalar $A$ and a pair of charged Higgs bosons $H^{\pm}$, while the complex scalar singlet serves as the DM candidate. This is equivalent to a two-component dark matter scenario. In the presence of a U(1) symmetry, the complex singlet scalar serves as the DM candidate while in the U(1) symmetry-breaking scenario the two dark matter candidates namely, the scalar and pseudoscalar components of the complex scalar singlet, are mass non-degenerate as discussed in sect. 4. Further, we focus on the dark matter phenomenology of 2HDMS consistent with all experimental constraints including dark matter, flavour and collider constraints in both U(1) symmetric and U(1) breaking scenarios. Dark matter candidates provide additional invisible decay modes for the Higgs bosons. Consequently associated production of the heavy Higgs bosons would lead to signatures involving visible SM quarks/leptons and missing energy signals. We choose a representative benchmark point to study collider prospects at the HL-LHC and at future $e^+e^-$ colliders and discuss the viability of observing the channel $2b+\not{E}_T$ channel at an $e^+e^-$ collider at $\sqrt{s}=3$ TeV. The salient points of this work include

- discussion of the parameter space allowed by theoretical and experimental constraints in 2HDMS;
- identification of some representative benchmarks for 2HDMS in both U(1) symmetric and U(1) breaking scenarios allowed by all experimental constraints including dark matter, flavour physics and collider constraints from the SM-like Higgs boson as well as from the heavy Higgs sector;
- collider phenomenology of the 2HDMS at the upcoming HL-LHC and at a high energy $e^+e^-$ collider.

The paper is organized as follows: we discuss the model followed by a discussion on the relevant theoretical and experimental constraints in sects. 2, 3. In sect. 4 we discuss the theoretical aspects of dark matter phenomenology for the model. In sect. 5 we study the U(1) symmetric case and in sect. 6 we study the U(1) breaking case. Our final results are summarized in Sect. 7.

## 2 The model

We consider the CP-conserving Type II two-Higgs-doublet model augmented with a complex scalar singlet (2HDMS) [16] to avoid flavour-changing neutral currents (FCNCs) at tree-level.[2] We consider soft $Z_2$-symmetry breaking consistent with tree-level FCNCs, allowing the presence of a mixing term between $\Phi_1$ and $\Phi_2$, i.e., $m_{12}^2$. The complex scalar singlet dark matter candidate $S$ is stabilized by a $Z_2'$ symmetry so that $S$ is odd under $Z_2'$ while the SM fields are even under the new $Z_2'$ symmetry as shown in Table 1. The $Z_2'$ symmetry remains unbroken both explicitly as well as dynamically, i.e., the singlet does not obtain a *vev*. The scalar potential $V_{2HDMS}$ with softly broken $Z_2$ and conserved $Z_2'$ symmetry is

$$V_{2HDMS} = V_{2HDM} + V_S \tag{1}$$

where the softly broken $Z_2$-symmetric 2HDM potential is

$$\begin{aligned} V_{2HDM} =\ & m_{11}^2\Phi_1^\dagger\Phi_1 + m_{22}^2\Phi_2^\dagger\Phi_2 - (m_{12}^2\Phi_1^\dagger\Phi_2 + h.c) + \frac{\lambda_1}{2}(\Phi_1^\dagger\Phi_1)^2 + \frac{\lambda_2}{2}(\Phi_2^\dagger\Phi_2)^2 + \\ & \lambda_3(\Phi_1^\dagger\Phi_1)(\Phi_2^\dagger\Phi_2) + \lambda_4(\Phi_1^\dagger\Phi_2)(\Phi_2^\dagger\Phi_1) + [\frac{\lambda_5}{2}(\Phi_1^\dagger\Phi_2)^2 + h.c] \end{aligned} \tag{2}$$

[1] Several interesting aspects also arise for cases where the singlet does develop a nonzero *vev* in 2HDMS [20].

[2] We have chosen the Type II 2HDM as it is a popular and widely used BSM theory, due to its connection to the Higgs sector of the Minimal Supersymmetric Standard Model (MSSM). However, note that besides Type II, there are other variations of 2HDM such as Type I and Type IV which also avoid FCNCs (for more details, see Ref. [30]).

**Table 1** The quantum numbers of the Higgs doublets and the singlet under the $Z_2$ and $Z_2'$ symmetry

| Fields | $Z_2$ | $Z_2'$ |
|---|---|---|
| $\Phi_1$ | +1 | +1 |
| $\Phi_2$ | −1 | +1 |
| $S$ | +1 | −1 |

and the $Z_2'$-symmetric singlet potential, $V_S$, is

$$
\begin{aligned}
V_S = {} & m_S^2 S^* S + \left(\frac{m_S^{2\prime}}{2} S^2 + h.c\right) \\
& + \left(\frac{\lambda_1''}{24} S^4 + h.c\right) + \left(\frac{\lambda_2''}{6}(S^2 S^* S) + h.c\right) + \frac{\lambda_3''}{4}(S^* S)^2 + S^* S[\lambda_1' \Phi_1^\dagger \Phi_1 + \lambda_2' \Phi_2^\dagger \Phi_2] \\
& + [S^2\left(\lambda_4' \Phi_1^\dagger \Phi_1 + \lambda_5' \Phi_2^\dagger \Phi_2\right) + h.c.].
\end{aligned}
\tag{3}
$$

The U(1) symmetric situation can be achieved for the case with $m_S^{2\prime}$, $\lambda_1''$, $\lambda_4'$, $\lambda_5'$=0. In this case, the complex scalar $S$ serves as a dark matter candidate as seen in sect. 2. In this case, the global U(1) symmetry stabilizes the dark matter due to a conserved quantum number such as a dark matter number assigned as an internal charge to the dark matter candidates as a consequence of the U(1) global symmetry [1].

In the U(1) breaking case, the $Z_2'$-symmetric potential $V_{2HDMS}$, eq. 1, allows for the U(1) symmetry breaking terms, involving $S^2$-, $S^4$-terms and the Hermitian conjugates.

In terms of its components, the Higgs doublets and singlet $S$ are parameterized as

$$\Phi_1 = (h_1^+ \quad \frac{1}{\sqrt{2}}(v_1 + h_1 + i a_1))^T, \tag{4}$$

$$\Phi_2 = (h_2^+ \quad \frac{1}{\sqrt{2}}(v_2 + h_2 + i a_2))^T, \tag{5}$$

$$S = \frac{1}{\sqrt{2}}(h_s + i a_s), \tag{6}$$

where $v_1$, $v_2$ are the vacuum expectation value(*vev*)'s obtained by the neutral components of $\Phi_1$ and $\Phi_2$, respectively. Since the singlet scalar does not obtain a *vev*, the minimization conditions are

$$m_{11}^2 = m_{12}^2 \frac{v_2}{v_1} - \lambda_1 \frac{v_1^2}{2} - \lambda_{345} \frac{v_2^2}{2}, \tag{7}$$

$$m_{22}^2 = m_{12}^2 \frac{v_1}{v_2} - \lambda_2 \frac{v_2^2}{2} - \lambda_{345} \frac{v_1^2}{2}, \tag{8}$$

where

$$\lambda_{345} = \lambda_3 + \lambda_4 + \lambda_5. \tag{9}$$

For simplicity, we made a choice $\lambda_1'' = \lambda_2''$. Although this does not affect collider phenomenology which is driven mainly by the portal couplings, $\lambda_1' - \lambda_5'$, it may affect dark matter phenomenology, the effect of which we do not focus in this work.

After electroweak symmetry breaking (EWSB), there remain 15 free parameters in the model.

$\lambda_1, \lambda_2, \lambda_3, \lambda_4, \lambda_5, m_{12}^2, \tan\beta, \lambda_1'', \lambda_3'', m_S^2, m_{S'}^2, \lambda_1', \lambda_2', \lambda_4', \lambda_5'$.

Here, $\tan\beta = \frac{v_2}{v_1}$ is the ratio of the *vev's* of the up-type and down-type Higgs doublet denoted by $v_2(= v \sin\beta)$ and $v_1(= v\cos\beta)$, respectively, where $v(= v_1^2 + v_2^2) \simeq 246$ GeV is the electroweak *vev*.

### 2.1 Fermion and gauge boson sector

For the Type II 2HDM, the up- and down-type quarks couple to the two different Higgs doublets. The down-type quarks and leptons couple to $\Phi_1$ while the up-type quarks couple to $\Phi_2$. Thus, the Yukawa Lagrangian is [30]

$$\mathcal{L}_{\text{Yukawa}} = y_u^{ij} Q_i \Phi_2 u_j - y_d^{ij} Q_i \Phi_1 d_j - y_l^{ij} L_i \Phi_1^{ij} l_j, \tag{10}$$

where $i$, $j = 1, 2, 3$ are the family indices of the fermions and $y_f$, $(f = u, d, l)$ are the Yukawa coupling matrices for the quarks and leptons. The couplings of the Higgs bosons to the quarks and leptons (normalized to the SM) are summarized in Table 2 [30].

The Higgs bosons couple to the gauge bosons as in the 2HDM [30], namely, the *HVV* and *HHVV* couplings are suppressed by $\cos(\beta - \alpha)$ as compared to the SM-like couplings of the Higgs boson ($\propto \sin(\beta - \alpha)$). For the CP-conserving scenario, the couplings of the pseudoscalar to two gauge bosons vanish at tree-level.

**Table 2** The fermion couplings in the Type II 2HDM [30]

| Particle | $h$ | $H$ |
|---|---|---|
| $u$ | $\cos\alpha/\sin\beta$ | $\sin\alpha/\sin\beta$ |
| $d$ | $-\sin\alpha/\sin\beta$ | $\cos\alpha/\cos\beta$ |
| $l$ | $-\sin\alpha/\sin\beta$ | $\cos\alpha/\cos\beta$ |

### 2.2 Higgs sector

In the absence of mixing between the Higgs doublets and the singlet, the Higgs sector of 2HDMS remains the same as in the 2HDM after electroweak symmetry breaking, consisting of two CP-even neutral scalar Higgs bosons $h$, $H$, a pseudoscalar Higgs $A$ and a pair of charged Higgs bosons $H^{\pm}$. The squared mass matrices for the charged ($\mathcal{M}^2_{\pm}$), the scalar ($\mathcal{M}^2_S$) and the pseudoscalar ($\mathcal{M}^2_{PS}$) Higgs sectors in the gauge eigenstate basis are given by [30],

$$M^2_{\pm} = [m^2_{12} - (\lambda_4 + \lambda_5)v_1 v_2]\begin{pmatrix} \frac{v_2}{v_1} & -1 \\ -1 & \frac{v_1}{v_2} \end{pmatrix}, \tag{11}$$

$$M^2_S = \begin{pmatrix} m^2_{12}\frac{v_2}{v_1} + \lambda_1 v_1^2 & -m^2_{12} + \lambda_{345} v_1 v_2 \\ -m^2_{12} + \lambda_{345} v_1 v_2 & m^2_{12}\frac{v_1}{v_2} + \lambda_2 v_2^2 \end{pmatrix}, \tag{12}$$

$$M^2_{PS} = \frac{m^2_A}{v_1^2 + v_2^2}\begin{pmatrix} v_2^2 & -v_1 v_2 \\ -v_1 v_2 & v_1^2 \end{pmatrix}, \tag{13}$$

where,

$$m^2_A = [\frac{m^2_{12}}{v_1 v_2} - 2\lambda_5](v_1^2 + v_2^2). \tag{14}$$

After EWSB, three Goldstone bosons are eaten up by the $W^{\pm}$ and $Z$ bosons while leaving two CP-even Higgses, a pseudoscalar and a pair of charged Higgs mass eigenstates.

### 2.3 Dark sector

The dark matter sector in 2HDMS consists of a variety of different possibilities depending on the symmetries involved. Referring closely to Ref. [1] which dealt with extending SM with a complex scalar singlet, similar cases may be studied for 2HDMS.

For conserved $Z_2'$ symmetry, $S \to -S$. In the component form, $h_s \to -h_s$ and $a_s \to -a_s$. In the absence of a vacuum expectation value ($v_s$) for both $h_s$ and $a_s$ constitutes the dark sector and interacts with the SM particles via the Higgs bosons. For the general case the squared masses of the two dark sector particles, are,

$$m^2_{h_s} = m^2_S + m^{2\prime}_S + (\lambda_1' + 2\lambda_4')\frac{v_1^2}{2} + (\lambda_2' + 2\lambda_5')\frac{v_2^2}{2} \tag{15}$$

$$m^2_{a_s} = m^2_S - m^{2\prime}_S + (\lambda_1' - 2\lambda_4')\frac{v_1^2}{2} + (\lambda_2' - 2\lambda_5')\frac{v_2^2}{2} \tag{16}$$

and the mass squared difference between $h_s$ and $a_s$ is,

$$\Delta^2 = |m^2_{h_s} - m^2_{a_s}| = 2(m^{2\prime}_S + \lambda_4' v_1^2 + \lambda_5' v_2^2) \tag{17}$$

We investigate the scenario for U(1) symmetric and breaking scenarios, respectively, below,

- U(1) symmetric case ($\lambda_1''$, $m^{2\prime}_S$, $\lambda_4'$, $\lambda_5' = 0$) In this case, $|\Delta| = 0$, hence both $h_s$ and $a_s$ components are mass-degenerate. This leads to a two component dark matter scenario with mass degenerate DM candidates. Alternately, it can be represented in terms of the complex scalar $S$ with a scalar and pseudoscalar component.
- U(1) breaking case ($\lambda_1''$, $m^{2\prime}_S$, $\lambda_4'$, $\lambda_5' \neq 0$) In this case, $|\Delta| > 0$ and there exists a nonzero mass difference between the $h_s$ and $a_s$ components.

As pointed out in Ref. [1] for SM + complex scalar singlet model, in the early universe, the annihilation of the heavier DM to the lighter DM is governed by the quartic couplings $\lambda_1''$, $\lambda_3''$ unless the mass difference between the two dark sector particles are small. We have checked using `micrOMEGAs` that both the dark sector particles are stable and there exists no decays for the heavier DM candidate to the lighter dark matter candidate at tree level for all benchmarks considered in our study.

Distinguishing between the single-component and two-component scenarios at the ILC has been recently studied in Ref. [31, 32], and similar interesting signals may also be relevant for 2HDMS but is beyond the scope of this work. We summarize the different possibilities we have discussed in Table. 3.

**Table 3** The different possibilities for DM in the U(1) symmetric and breaking scenarios

| | $U(1)$ | $U(1)$ breaking |
|---|---|---|
| Free parameters | $m_S^{2'}, \lambda_1'', \lambda_4', \lambda_5' = 0$ | $m_S^{2'}, \lambda_1'', \lambda_4', \lambda_5' \neq 0$ |
| Mass of DM | $m_{h_s} = m_{a_s}$ | $m_{h_s} \neq m_{a_s}$ |
| DM candidates | $S$ | $h_s, a_s$ |

### 2.4 Complex singlet scalar potential in terms of two real scalars

In this subsection, we discuss an alternate representation of the complex scalar singlet potential in terms of its real components $h_s$ and $a_s$. The squared masses of the mass eigenstates in the dark sector are shown in eqs. 15 and 16.

The interaction terms in the singlet scalar potential are

$$V_{int} = \frac{1}{48}[\lambda_1''(h_s^2 - a_s^2)^2 + 3\lambda_3''(h_s^2 + a_s^2)^2 + 4\lambda_2''(h_s^4 - a_s^4)]. \tag{18}$$

The singlet components interact with the Higgs bosons via the trilinear and the quartic interaction terms at tree-level. The —for our discussion relevant—trilinear couplings are:

$$\lambda_{hh_sh_s} = 2v((\lambda_1' + 2\lambda_4')\cos\beta\sin\alpha - (\lambda_2' + 2\lambda_5')\sin\beta\cos\alpha), \tag{19}$$

$$\lambda_{Hh_sh_s} = -2v((\lambda_1' + 2\lambda_4')\cos\beta\cos\alpha + (\lambda_2' + 2\lambda_5')\sin\beta\sin\alpha), \tag{20}$$

$$\lambda_{ha_sa_s} = 2v((\lambda_1' - 2\lambda_4')\cos\beta\sin\alpha - (\lambda_2' - 2\lambda_5')\sin\beta\cos\alpha), \tag{21}$$

$$\lambda_{Ha_sa_s} = -2v((\lambda_1' - 2\lambda_4')\cos\beta\cos\alpha + (\lambda_2' - 2\lambda_5')\sin\beta\sin\alpha), \tag{22}$$

### 2.5 Discussions

In the U(1) limit, the squared masses of the mass eigenstates in the dark sector reduce to

$$m_{h_s}^2 = (m_S^2 + \lambda_1'\frac{v_1^2}{2} + \lambda_2'\frac{v_2^2}{2}), \tag{23}$$

$$m_{a_s}^2 = (m_S^2 + \lambda_1'\frac{v_1^2}{2} + \lambda_2'\frac{v_2^2}{2}), \tag{24}$$

and the mass squared difference between $h_s$ and $a_s$ is,

$$\Delta^2 = |m_{h_s}^2 - m_{a_s}^2| = 0 \tag{25}$$

due to the mass degenerate particles. Therefore, $h_s$ and $a_s$ are mass degenerate. The relevant trilinear couplings are:

$$\lambda_{hh_sh_s} = 2v(\lambda_1'\cos\beta\sin\alpha - \lambda_2'\sin\beta\cos\alpha), \tag{26}$$

$$\lambda_{Hh_sh_s} = -2v(\lambda_1'\cos\beta\cos\alpha + \lambda_2'\sin\beta\sin\alpha), \tag{27}$$

$$\lambda_{ha_sa_s} = 2v(\lambda_1'\cos\beta\sin\alpha - \lambda_2'\sin\beta\cos\alpha), \tag{28}$$

$$\lambda_{Ha_sa_s} = -2v(\lambda_1'\cos\beta\cos\alpha + \lambda_2'\sin\beta\sin\alpha), \tag{29}$$

For the rest of the study, we first discuss the features of the DM observables for the U(1) symmetric case followed by the U(1) breaking case. Before moving on to phenomenology, let us first discuss the important constraints to which the model is subjected to.

## 3 Constraints

### 3.1 Theoretical constraints

#### *3.1.1 Boundedness-from-below conditions*

Boundedness-from-below (BFB) of the scalar potential ensures that $V_{2HDMS} \geq 0$. We discuss first the conditions for the U(1) symmetric case as below.

For solving for the directions involving $h_1$, $h_2$, $h_s$, we follow the criteria of copositivity and Sylvester's criterion as defined in Ref.[33], one first obtains the biquadratic matrix $\mathcal{Q}$ in the eigenbasis ($h_1^2\ h_2^2\ h_s^2$) for minima,

$$\mathcal{Q} = \frac{1}{8}\begin{pmatrix} \lambda_1 & \lambda_3+\lambda_4-|\lambda_5| & \lambda_1' \\ \lambda_3+\lambda_4-|\lambda_5| & \lambda_2 & \lambda_2' \\ \lambda_1' & \lambda_2' & 2\lambda_3'' \end{pmatrix} \tag{30}$$

Using copositivity and Sylvester criterion [33], one obtains the necessary and sufficient boundedness from below conditions as,

$$\begin{aligned} \lambda_1 &\geq 0, \\ \lambda_2 &\geq 0, \\ \lambda_3+\lambda_4-|\lambda_5|+\sqrt{\lambda_1\lambda_2} &\geq 0, \\ \lambda_3'' &\geq 0, \\ \lambda_1' + \sqrt{\frac{1}{2}\lambda_1\lambda_3''} &\geq 0, \\ \lambda_2' + \sqrt{\frac{1}{2}\lambda_2\lambda_3''} &\geq 0 \\ \sqrt{2\lambda_1\lambda_2\lambda_3''} + \lambda_2\sqrt{2\lambda_3''} & \\ +\lambda_2'\sqrt{\lambda_1} + \sqrt{2(\lambda_3+\lambda_4-\lambda_5+\sqrt{\lambda_1\lambda_2})} & \\ \sqrt{(\lambda_1'+\sqrt{2\lambda_2\lambda_3''})(\lambda_2'+\sqrt{2\lambda_1\lambda_3''})} &\geq 0 \end{aligned} \tag{31}$$

For the U(1) breaking case, due to the multiple possible scalar directions and a vast number of terms in the potential, the analytical computation of the BFB conditions is rather complicated. We follow the numerical procedure following Ref. [33] and derived numerically for 2HDMS in Ref. [25].

#### *3.1.2 Vacuum stability*

We have ensured all the benchmarks considered in the study satisfy vacuum stability conditions numerically using `EVADE`[34–36] to ensure that no deeper minima exists than the one we consider with $v_s = 0$ where $v_s$ is the vacuum expectation value (vev) obtained by the real scalar component $h_s$ of the complex scalar singlet. The imaginary component of the complex singlet is assumed to not obtain a vev to avoid CP-violation. This ensures that the complex scalar singlet does not obtain a vev and acts as a two-component dark matter candidate for the rest of the study.

### 3.2 Experimental constraints

The following experimental constraints are imperative for any BSM model.

- The lightest CP-even Higgs mass ($m_h$) is measured as 125.25±0.17 GeV within the experimental error [37].
- The invisible decay width of the light Higgs boson is constrained by ATLAS and CMS as given below,

$$\begin{aligned} BR(h \to \chi\chi) &\leq 0.11^{+0.04}_{-0.03}\ \text{(ATLAS) [38]} \\ &\leq 0.19\ \text{(CMS)[39]}. \end{aligned} \tag{32}$$

This limit is adhered to for the choice of $m_{DM} > 62.5$ GeV.

- Flavour physics constraints, as taken into account, namely $BR(b \to s\gamma) = (3.55\pm0.24\pm0.09)\times10^{-4}$ [40], $BR(B_s \to \mu^+\mu^-) = (3.2^{+1.4+0.5}_{-1.2-0.3}) \times 10^{-9}$ [41, 42]. The benchmark points also respect the upper limit of $\Delta(g-2)_\mu (= 261(63)(48) \times 10^{-11})$ [43]. We also ensure that the benchmarks are within the SM expectation values of $\epsilon_K$[44], $\Delta M_{B_q}$ (where $q = s, d$)[45].
- The benchmark points satisfy the electroweak precision test constraints on the STU parameters, where S = 0.02 ± 0.1, T = 0.07±0.12, U = 0.00±0.09 [46].
- The relic density upper limit is adhered to from `PLANCK` data, i.e., $\Omega h^2 = 0.11933 \pm 0.00091$ [47].
- The DM-nucleon spin-independent cross sections satisfy constraints from `LUX-ZEPLIN` (LZ) [6] and the indirect detection constraints from `Fermi-LAT` [48, 49].
- Collider constraints from LEP [50] and Run 2 ATLAS/CMS searches on the heavy Higgs searches [51, 52] and the 125 GeV Higgs signal strength measurements [53] have been taken into account.

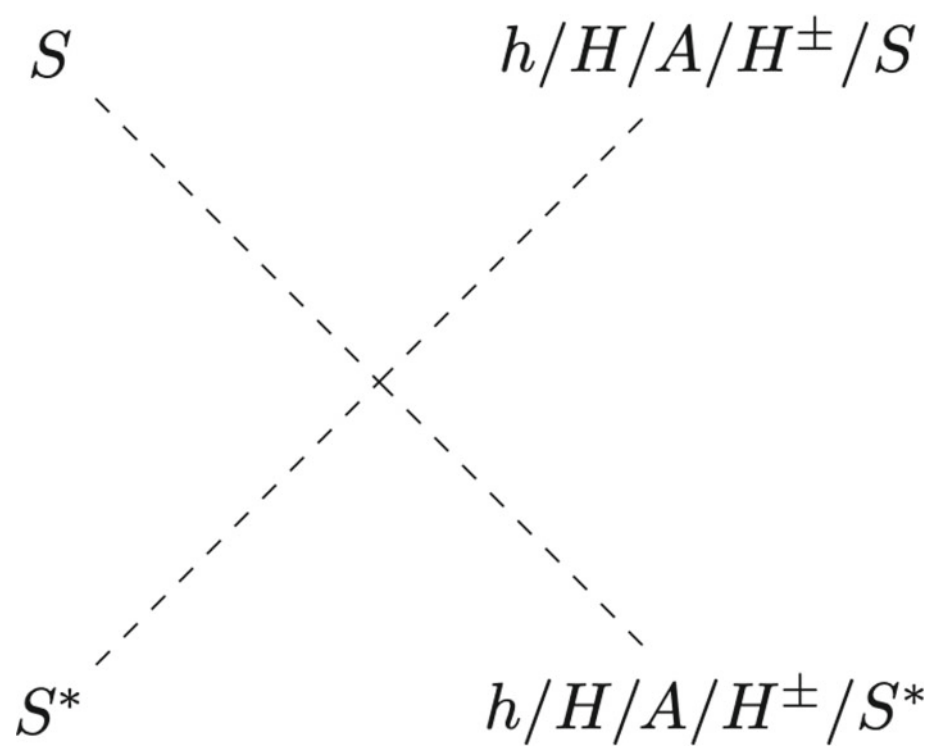

**Fig. 1** The four-point vertices contribute to the computation of the relic density

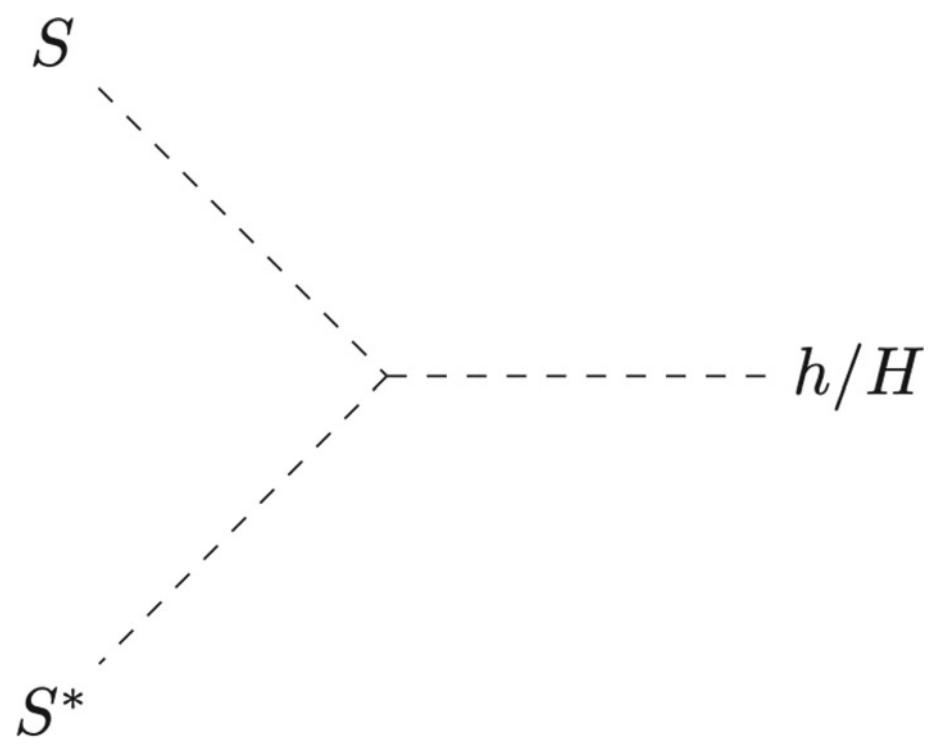

**Fig. 2** The vertex contributing to the *s*-channel Higgs mediated processes contributing to the relic density. The Higgs bosons decay to all possible final states including fermions, gluons, gauge bosons as well as lighter Higgs bosons arising from the decay of the heavier CP-even Higgs

The Higgs sector of the 2HDMS is the same as in the 2HDM (for the current status of the Higgs sector, see [54, 55]). For 2HDMS, the model is implemented using `SARAH-v4.14.3` [56] and `SPheno-4.0.4` [57] is used for generating the particle mass spectra and decays, and for performing the parameter scans. `SARAH` provides the CalcHEP files required for computing the DM relic and direct detection constraints using `micrOMEGAs` [58]. Constraints on the Higgs sector are checked using `HiggsBounds` and `HiggsSignals` [59]. Constraints from B-physics and other flavour observables are checked using `SPheno-v4.0.4` [57].

## 4 Dark matter phenomenology

In the absence of a vacuum expectation value, the complex scalar singlet acts as a dark matter candidate. It interacts with the SM particles via the Higgs bosons which act as scalar mediators. For both single and multi-component scenarios, no spin-dependent interactions occur for the dark matter candidate and the only stringent constraints arise from the spin-independent DM-nucleon direct detection searches. We discuss briefly the theoretical framework for the computation of the relic density and direct detection cross sections.

### 4.1 Relic density

For thermal dark matter, i.e., where the DM and SM are in thermal equilibrium, the processes contributing to relic density are the annihilation and the co-annihilation processes of the DM with the SM particles, see Fig. 1-3. Besides self-interactions, the only interaction of the DM candidate to the SM particles is via the Higgs bosons (at tree-level), only s-channel Higgs $h(H)$ mediated channels as shown in Fig. 2 contribute besides t-channel contributions mediated via $S$.

### 4.2 Spin-independent direct detection cross section

The interaction of the DM with nucleon is mediated by the t-channel neutral scalar Higgs bosons at tree-level. The relevant process is shown in Fig. 4. The vertex involved in this processes is $\lambda_{h_i SS^*}$.

The direct detection cross section rate for the proton and neutron scattering is

$$\sigma_p^{SI} = \frac{4\mu_p^2}{\pi}[f_p Z]^2, \tag{33}$$

$$\sigma_n^{SI} = \frac{4\mu_n^2}{\pi}[f_n(A-Z)]^2, \tag{34}$$

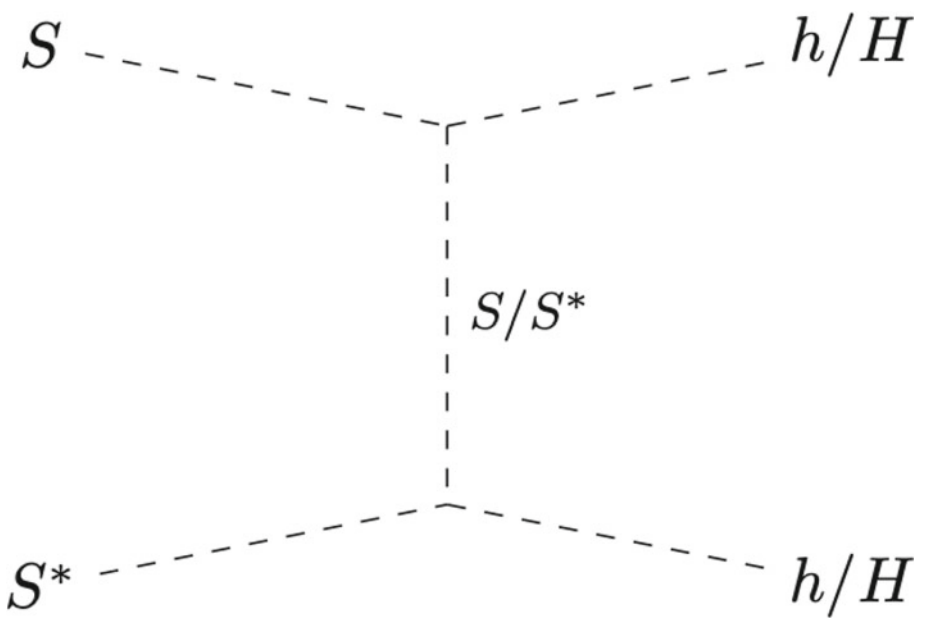

**Fig. 3** The $t$-channel mediated processes that contribute to the relic density

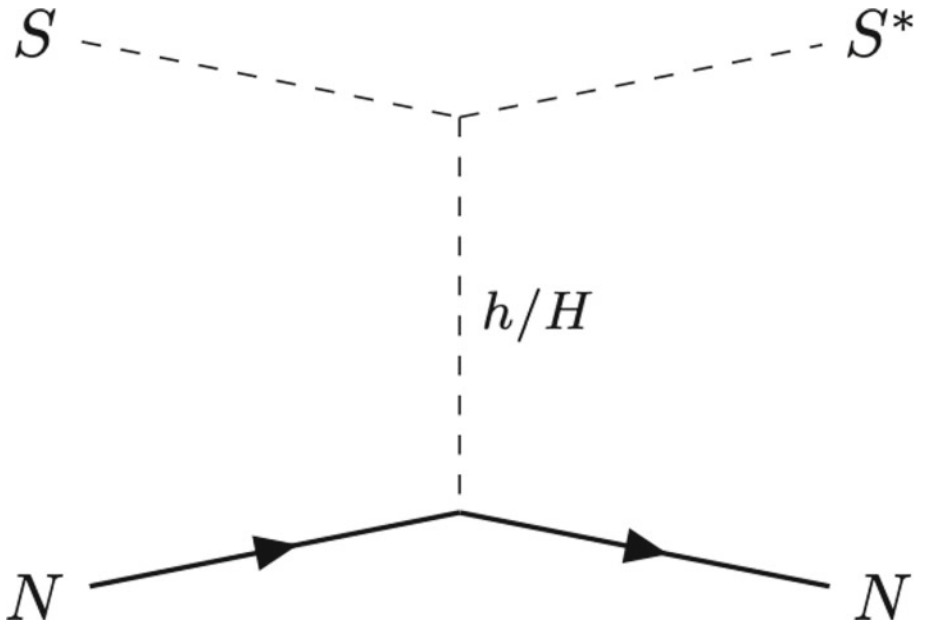

**Fig. 4** The t-channel process corresponding to direct detection of the dark matter candidate $S$. Here, $N$ refers to the nucleon (proton or neutron)

where

$$\mu_{p,n} = \frac{m_{p,n} m_\chi}{m_{p,n} + m_\chi} \tag{35}$$

is the reduced mass of the DM-nucleon system where $p$, $n$ refer to the proton and neutron, respectively, and $m_\chi$ is the mass of the dark matter candidate. The couplings of the proton $f_p$ and neutron $f_n$ and to the DM are computed from the DM-quark scattering amplitudes, see eq. 36. The heavy Higgs-mediated process would be relatively suppressed compared to the SM Higgs-mediated process due to $m_H > m_h$. The DM-quark amplitude $\mathcal{M}$ is,

$$\mathcal{M} = \sum_i \frac{\lambda_{h_i SS^*} \lambda_{h_i q\bar{q}}}{t - m_{h_i}^2} \tag{36}$$

where $i = 1, 2$ such that $h_i = h,\ H$ and $h_i q\bar{q}$ are the Yukawa couplings of the quarks to the Higgs bosons. Assuming DM number conservation and for zero momentum transfer, the propagator reduces to $\frac{1}{-m_{h_i}^2}$.

### 4.3 Indirect detection

Dark matter self-annihilation and the interactions via the CP-even Higgs bosons and their decays into final state SM particles lead to indirect detection signatures. Among these signatures, stringent constraints arise from Fermi-LAT [48, 49] on the channels $b\bar{b}$ and $\tau\tau$ followed by $W^+W^-$. In Higgs portal models, the annihilation channel in the final state $b\bar{b}$ is crucial for indirect detection, as $b\bar{b}$ constitutes the dominant decay of the SM-like Higgs boson. We have checked using `micrOmegas−v5.2.4` to ensure the benchmark is allowed by the current Fermi-LAT data.

## 5 The U(1) symmetric case

In this section, we consider the complex scalar singlet as the DM candidate for the U(1) symmetric case.

### 5.1 Relevant couplings

The scalar Higgs bosons couple to the dark matter via trilinear and quartic interactions. The relevant trilinear couplings of the DM with the CP-even Higgs bosons are

$$\lambda_{hS^*S} = \frac{v}{\sqrt{1+\tan^2\beta}}(\lambda_1' \sin\alpha - \lambda_2' \tan\beta \cos\alpha), \tag{37}$$

$$\lambda_{HS^*S} = \frac{-v}{\sqrt{1+\tan^2\beta}}(\lambda_1' \cos\alpha + \lambda_2' \tan\beta \sin\alpha). \tag{38}$$

**Table 4** Relevant parameters of the benchmark **BP1** used for the study. All mass parameters have units GeV except for $m_{12}^2$ in GeV$^2$. The decimal points are rounded to the first decimal place for the masses and up to the third decimal place for the relic density, and the direct detection cross sections

| Parameters | **BP1** |
|---|---|
| $\lambda_1$ | 0.23 |
| $\lambda_2$ | 0.26 |
| $\lambda_3$ | 0.2 |
| $\lambda_4$ | −0.14 |
| $\lambda_5$ | 0.10 |
| $m_{12}^2(\text{GeV}^2)$ | $1.0\times10^5$ |
| $\lambda_3''$ | 0.1 |
| $\lambda_1'$ | 2.0 |
| $\lambda_2'$ | 0.01 |
| $\tan\beta$ | 6.8 |
| $\cos(\alpha-\beta)$ | −0.00078 |
| $m_h$ (GeV) | 125.16 |
| $m_H$(GeV) | 833.54 |
| $m_A$(GeV) | 829.84 |
| $m_{H^\pm}$ (GeV) | 833.49 |
| $m_\chi$(GeV) | 401.97 |
| $\Omega h^2$ | $3.56\times10^{-3}$ |
| $\sigma_p^{SI} * \frac{\Omega h^2}{\Omega h^2{}_{PLANCK}} \times 10^{10}$ (pb) | 0.048 |
| $\sigma_n^{SI} * \frac{\Omega h^2}{\Omega h^2{}_{PLANCK}} \times 10^{10}$ (pb) | 0.051 |

**Table 5** Range of the relevant parameter varied for the scans involving the relic density and the direct detection cross section versus the dark matter mass, $m_\chi$. The other parameters, $\lambda_1 = 0.23$, $\lambda_2 = 0.26$, $\lambda_3 = 0.20$, $\lambda_4 = -0.14$, $\lambda_5 = 0.10$, $m_{12}^2 = 10^5$, $\tan\beta = 6.8$, $\lambda_3'' = 0.1$, $\lambda_1' = 2.0$ and $\lambda_2' = 0.01$ are kept fixed throughout the scan

| Parameters | $m_S^2$ (GeV$^2$) |
|---|---|
| Values | 0-1000000 |

### 5.2 Parameter scans

In this section, we scan over the parameter space starting from a chosen representative benchmark **BP1** (as shown in Table 4) for 2HDMS consistent with all theoretical constraints and experimental constraints from the Higgs, flavour, collider and dark matter constraints. The tree-level unitarity constraints are checked using `SPheno-4.0.4`. We ensure that the benchmark satisfies the boundedness-from-below conditions. We ensure the benchmark passes all theoretical and experimental constraints.

We scan over the variable $m_S^2$ in the range in Table. 5 to study the variation of the DM observables with the dark matter mass.

### 5.3 Relic density

In Fig. 5, we plot the relic density against the DM mass. We observe a funnel region near $m_\chi \simeq m_h/2$ where the dark matter resonant annihilation occurs via the lightest CP-even Higgs with a mass of 125 GeV. The dominant processes contributing to the relic density in this region are $b\bar{b}$ and sub-dominant contributions from $WW$. As one moves away from the funnel region, the $WW$ mode starts dominating over $b\bar{b}$. This is due to the increasing phase space available to the final state particles due to the increase in mass of the dark matter candidate in the initial states. This continues up to masses close to $m_W$ while near $m_\chi \simeq m_W$, the dominating mode is $WW$. As one increases $m_\chi$ further the $ZZ$ mode is now allowed once the $m_\chi \simeq m_Z$. In this region, both the $WW$ and $ZZ$ modes contribute to the relic density computation. As $m_\chi \simeq m_h$, the di-Higgs channel opens up. Further, for $m_\chi$ near the top mass, the $t\bar{t}$ channel opens up. For higher DM masses, the relic density is mostly overabundant except for $m_\chi \simeq m_H/2$ where $t\bar{t}$ and $b\bar{b}$ modes dominate. Near the coannihilation regions, where $m_\chi \simeq m_H$, heavy Higgs channels open up such as $H^+H^-$, $hH$, $AZ$, $H^\pm W^\mp$ and $HH$ followed by sub dominant contributions from SM modes. As one moves away from the heavy Higgs resonance region and the coannihilation region, the dominant channels contributing to the relic density is $H^+H^-$, $AA$ and $HH$ with negligible contributions from other di-Higgs channels and SM modes.

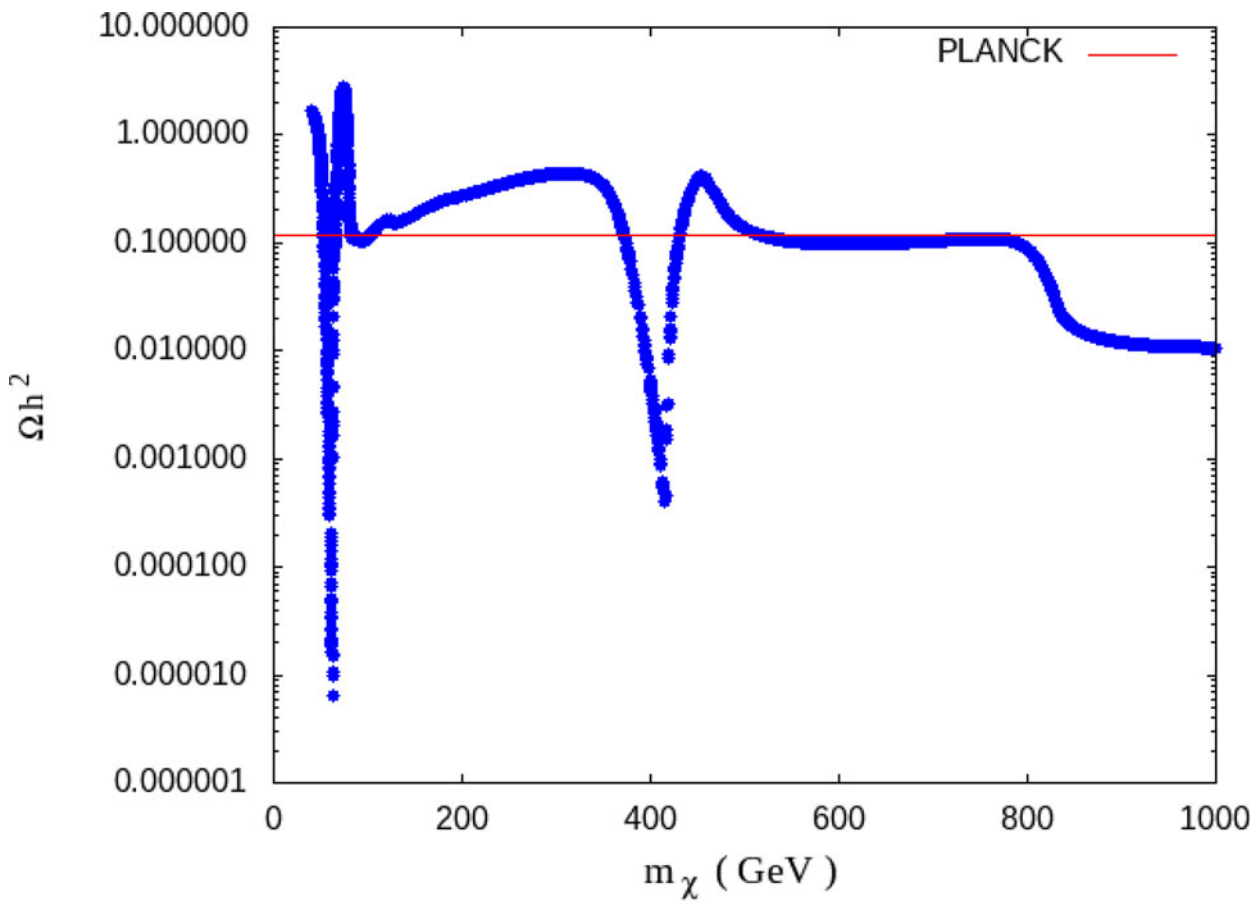

**Fig. 5** Variation of the relic density vs. mass of dark matter $m_\chi$. The parameter $m_S^2$ is varied in this plot in the range given in Table 5. The other parameters, $\lambda_1 = 0.23$, $\lambda_2 = 0.26$, $\lambda_3 = 0.20$, $\lambda_4 = -0.14$, $\lambda_5 = 0.10$, $m_{12}^2 = 10^5$, $\tan\beta = 6.8$, $\lambda_3'' = 0.1$, $\lambda_1' = 2.0$, and $\lambda_2' = 0.01$ are kept fixed throughout the scan

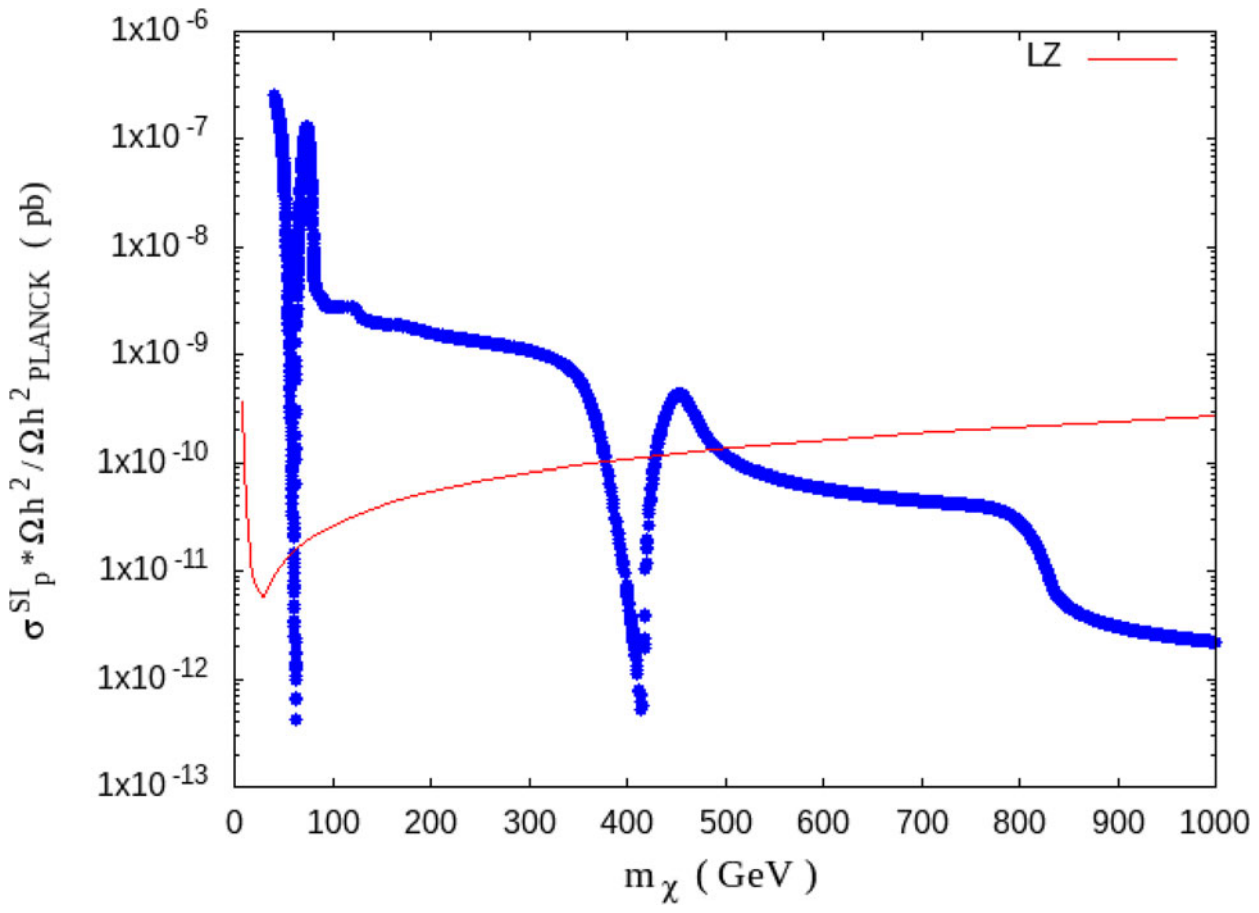

**Fig. 6** Variation of the spin independent direct detection cross section against the mass of the dark matter candidate $m_\chi$. The parameter $m_S^2$ is varied in this plot in the range given in Table 5. The other parameters, $\lambda_1 = 0.21$, $\lambda_2 = 0.26$, $\lambda_3 = 0.199$, $\lambda_4 = -0.10$, $\lambda_5 = 0.10$, $m_{12}^2 = 10^5$, $\tan\beta = 6.8$, $\lambda_3'' = 0.1$, $\lambda_1' = 2.0$, and $\lambda_2' = 0.01$ are kept fixed throughout the scan

Therefore, in summary, from Fig. 5, we observe that the relic density remains underabundant in the resonance regions corresponding to $m_\chi \sim m_{h/2}$ and $m_H/2$ while thermal relic is obtained close to the coannihilation region to $m_\chi \sim m_H$. The relic density as a function of dark matter mass decreases for higher dark matter masses due to the couplings of the dark matter candidate (with the exception of the resonance region). Further, we have checked for different values of the portal couplings that similar features of resonant annihilations are observed and are generic predictions of the model.

### 5.4 Direct detection

The relevant constraints from direct detection cross section of dark matter arises from the spin-independent (SI) interactions. Figure 6 shows the variation of the scaled direct detection cross section against the dark matter mass $m_\chi$. In the event there are other sources of dark matter besides the complex singlet as requisite in this case to fulfill the relic density, the direct detection cross section from LZ by scaling it with the ratio of the relic density in 2HDMS to the PLANCK result as must be the case for multi-component dark matter. This is because the limits from LUX ZEPLIN (LZ) assume thermal relic and single component dark matter. For multi-component DM studies, the direct detection cross section must be scaled by a factor $\frac{\Omega h^2}{\Omega h^2_{PLANCK}}$ (as also shown in some references in Ref. [1, 60]).

From Fig. 6, we observe that the dark matter direct detection stringently rules out dark matter masses below the heavy Higgs resonance region. For the coannihilation region, i.e., $m_\chi \sim m_H$, one has the possibility of heavy dark matter satisfying both thermal relic and direct detection cross section. Such points with $m_\chi \sim m_H$, however, do not lead to any invisible branching of the heavy Higgs $H$ therefore with no prospects for any collider signal.

### 5.5 Collider analysis

In this section, we discuss the signatures of this model at the HL-LHC and at future $e^+e^-$ colliders. As discussed in the previous section, the presence of the invisible decay of the Higgs bosons to the dark matter candidate is a source of missing energy signals at the colliders. Therefore, the direct production of heavy Higgs bosons and the consecutive decay of the heavy Higgs bosons into $\chi$ visible SM particles can give rise to distinct signatures for this scenario as compared to the 2HDM-like scenario. We investigate the collider prospects of 2HDMS in the context of both $\sqrt{s}=14$ TeV LHC at the targeted integrated luminosity of 3-4 ab$^{-1}$ and in future $e^+e^-$ colliders up to $\sqrt{s}=3$ TeV and integrated luminosities 5 ab$^{-1}$.

**Table 6** Relevant parameters of the benchmarks used for the study. All mass parameters have units GeV except for $m_{12}^2$ in GeV$^2$. The decimal points are rounded to the first decimal place for the masses and up to the third decimal place for the relic density, and the direct detection cross sections

| Parameters | **BP1** | **BP2** |
|---|---|---|
| $\lambda_1$ | 0.21 | 0.21 |
| $\lambda_2$ | 0.26 | 0.26 |
| $\lambda_3$ | 0.2 | 0.2 |
| $\lambda_4$ | $-0.1$ | $-0.1$ |
| $\lambda_5$ | 0.1 | 0.1 |
| $m_{12}^2$(GeV$^2$) | $1.0\times10^5$ | $1.0\times10^5$ |
| $\lambda_3''$ | 0.1 | 0.1 |
| $\lambda_1'$ | 2.0 | 3.0 |
| $\lambda_2'$ | 0.01 | 0.01 |
| $\tan\beta$ | 6.8 | 6.8 |
| $\cos(\alpha-\beta)$ | -0.00078 | $-0.00078$ |
| $m_h$ (GeV) | 125.16 | 125.16 |
| $m_H$(GeV) | 833.54 | 833.54 |
| $m_A$(GeV) | 829.84 | 829.84 |
| $m_{H^\pm}$ (GeV) | 833.49 | 833.49 |
| $m_\chi$(GeV) | 401.97 | 402.77 |
| $\Omega h^2$ | $3.56\times10^{-3}$ | $1.52\times10^{-3}$ |
| $\sigma_p^{SI} * \frac{\Omega h^2}{\Omega h^2{}_{PLANCK}} \times 10^{10}$ ( pb) | 0.048 | 0.0021 |
| $\sigma_n^{SI} * \frac{\Omega h^2}{\Omega h^2{}_{PLANCK}} \times 10^{10}$ ( pb) | 0.051 | 0.0023 |
| BR($H\to\chi\chi$) ($\times10^{-2}$) | 1.94 | 4.67 |

**Table 7** The branching ratios for the dominant decay modes of the heavy Higgs bosons for **BP1**. The branching ratios are rounded up to the third decimal place

| Decay Channels | Branching ratios for **BP1** |
|---|---|
| $H\to b\bar{b}$ | 0.326 |
| $H\to t\bar{t}$ | 0.603 |
| $H\to\tau\bar{\tau}$ | 0.050 |
| $H\to\chi\bar{\chi}$ | 0.019 |
| $A\to b\bar{b}$ | 0.304 |
| $A\to t\bar{t}$ | 0.648 |
| $A\to\tau\bar{\tau}$ | 0.046 |
| $H^\pm\to t\bar{b}$ | 0.962 |
| $H^\pm\to\tau\bar{\nu_\tau}$ | 0.035 |

**BP1** is a heavy DM benchmark point as seen in Table 6 where the dark matter mass$\sim$ 402 GeV and is allowed by direct detection data. The dominant decay modes of the heavy Higgs bosons are summarized in Table 7. **BP2** represents a similar benchmark as **BP1** with a higher branching ratio of the heavy Higgs boson to dark matter. In the presence of the dark matter, there are additional decay channels opening up for the heavy Higgs, $H\to\chi\chi$. The presence of the invisible DM candidate in the final state ensures the presence of missing energy in the final state signal at colliders. Such signatures have been recently studied with the heavy Higgs as a portal to dark matter and its collider signals of mono-jets and VBF along with missing transverse energy at the LHC [15].

### *5.5.1 Simulation details*

We use `MG5_aMC_v3_1 _1` [61, 62] to generate the parton level hard scattering processes for both signal and SM background. Hadronization and showering of the parton level events is performed using `Pythia-8` [63] while a fast detector simulation is performed using `Delphes-v3.4.1` [64–66] with the default Delphes card provided for ATLAS for the LHC study. For the $e^+e^-$ study, we used the default Delphes card for the ILD detector based on [67]. We performed the signal-to-background analysis for LHC using `Delphes-v3.4.1` and `Madanalysis-v5` [68–72] for the electron-positron collider.

### 5.5.2 Prospects at LHC

The main processes contributing to neutral heavy Higgs production are gluon fusion (ggF)(dominantly mediated by the top quark loop followed by the bottom quark loop), vector boson fusion (VBF) and associated Higgs production $(Vh_i)$, $b\bar{b}h_i$, $t\bar{t}h_i$ [30]. For the charged Higgs pair, the possible production channels are $H^+H^-$ and $W^{\pm}H^{\mp}$ [30]. The current constraints on heavy Higgs bosons are summarized in [51, 52]. Projection studies of heavy Higgs bosons predict a mass reach of up to TeV for direct searches [73, 74] and indirect searches [75]. The HL-LHC is important to achieve the required luminosity to observe these channels and gain indirect insight into the BSM Higgs sector. The VBF and gluon fusion channel have been studied for the real singlet at LHC and are good discovery probes [15]. For the mass ranges of the heavy Higgs bosons considered in our study, the dominant production processes at $\sqrt{s} = 14$ TeV LHC are: $b\bar{b}H$, $H + jj(ggF)$, $VBF$, $ZH$ and $t\bar{t}H$. The *WH* associated production is suppressed due to the suppressed couplings of the heavy Higgs bosons with respect to the SM couplings to the gauge bosons. The production cross sections for $b\bar{b}H$ and $t\bar{t}H$ are also substantially suppressed compared to ggF and VBF. Therefore, the most sensitive probes for neutral heavy Higgs production are ggF and VBF.

In the presence of the heavy Higgs $H$ decaying to two dark matter candidates, one can obtain invisible momentum signatures in the final state. Keeping this in mind, one can look into the following final states:

- $1j$ (ISR)+$\not{E}_T$ [76]
- $2j + \not{E}_T$ [77]

We estimate the significance for the mono-jet and VBF channels using the cuts from an existing cut-and-count analyses performed in Ref. [15] for $\sqrt{s} = 14$ TeV LHC.

### 5.5.3 Pre-selection cuts

The pre-selection cuts used are summarized below. We choose the default Delphes ATLAS card.

- The leptons are reconstructed with a minimum transverse momentum, $p_T > 10$ GeV and pseudorapidity $|\eta| < 2.5$ while excluding the transitional pseudorapidity gap between the barrel and the end cap of the calorimeter $1.37 < |\eta| < 1.52$.
- Photons are reconstructed with $p_T > 10$ GeV and $|\eta| < 2.5$.
- All jets are reconstructed using $\Delta R = 0.4$ using the anti-$k_T$ algorithm and minimum $p_T > 20$ GeV with pseudorapidity $|\eta| < 2.5$.

### 5.5.4 Signal region A: $1j + \not{E}_T$

We estimate the signal events for the monojet $+\not{E}_T$ channel using the SM background estimates in Ref. [15]. Using the kinematic cuts, at least four jets, $\Delta\Phi(j, \not{E}_T) > 0.4$, $p_T(j) > 250$ GeV and $\not{E}_T > 250$ GeV and lepton vetoes, we obtain a signal significance of $6.19\sigma$ for **BP1** and $14.67\sigma$ for **BP2** at 3 ab$^{-1}$ using gluon fusion production channel at NNLO+NNLL.

### 5.5.5 Signal region B: $2j + \not{E}_T$

We study the VBF topology for the signal region consisting of two forward jets and $\not{E}_T$. We follow the cuts in Ref. [15] for estimating the signal efficiency and use the background estimates from the paper. Owing to the very low production cross section and low invisible branching ratio of the heavy CP-even Higgs, there is no sensitivity in this channel at HL-LHC for **BP1** and **BP2**.

Therefore, we observe that owing to the small invisible branching ratio and heavy Higgs masses $\sim 830$ GeV (and hence small production cross section) in **BP1**, and the final states are inaccessible at the upcoming HL-LHC run. New machine learning techniques have shown an improvement for real singlet extended 2HDM as in Ref. [15] and may also be beneficial for the 2HDMS but is beyond the scope of our current work. Taking a different direction, we compare the reach in an e$^+$ e$^-$ collider and estimate the prospects of observing such a benchmark at the $e^+e^-$ collider using a cut-and-count approach.

### 5.5.6 Prospects at e$^+$e$^-$ colliders

The cleaner environment compared to hadron colliders make the electron-positron linear collider attractive for precision studies of new physics. The International Linear Collider (ILC) [78] is a proposed $e^+e^-$ collider with center-of-mass energies at the SM-like Higgs threshold ($\sqrt{s} = 250$ GeV), top threshold ($\sqrt{s} = 350$ GeV) and with further upgrades up to center of mass energies of $\sqrt{s} = 500$ GeV up to $\sqrt{s} = 1$ TeV with an annual luminosity of about $\mathcal{L} = 500$ fb$^{-1}$. The ILC offers both the possibility of exploiting the polarization of the electron and positron beams leading to an increased background suppression, higher sensitivity to specific couplings and further new observables offering a rich window for unique phenomenology [79]. Other proposed $e^+e^-$ colliders are CLIC [80, 81] with an energy upgrade up to $\sqrt{s} = 3$ TeV, FCC-ee [82] and CEPC [83] with the latter having beam energies up to the $t\bar{t}$ threshold.

As a representative study, we perform a signal-to-background analysis at a generic $e^+e^-$ collider with $\sqrt{s} = 3$ TeV and unpolarized electrons and positrons to estimate the observability of these channels after background rejection for **BP1** and **BP2**. The presence of invisible dark matter in the final state manifests as missing energy in the final state along with visible SM-particles. For a detailed assessment, the inclusion of the polarized beams and for different luminosity running scenarios is required, which is postponed to a forthcoming study.

#### *5.5.7 Signals and backgrounds*

From Table 7, we observe that the dominant decay of the pseudoscalar is to a pair of top quarks and $b$ quarks. For **BP1** a small fraction of the CP-even heavy Higgs decays to dark matter which manifests itself as missing energy at the collider. In such cases. the dominant production processes $ZH$,$HA$ and off-shell contributions to $b\bar{b}H$, $t\bar{t}H$.

This leads to a final state including two b-jets and missing energy while $Hjj$ leads to a final state with two light jets associated with the missing energy with the latter having a VBF topology. On the other hand, the production channel $\nu\bar{\nu}H$ state leads to a fully invisible state which in order to be probed requires an ISR photon against which the invisible system recoils. Besides, note that the heavy Higgs bosons ($A$) also decay to a pair of tau leptons, therefore allowing the final state of $2\tau + \not{E}_T$ from the $HA$ production channel. Since the branching ratio of the $A$ is dominantly into $t\bar{t}$ followed by $b\bar{b}$, one expects final states involving at least $2b + \not{E}_T$ to be dominant over the tau final states. In this work, we look into the prospects of observing signals including $2b + \not{E}_T$. Thus, the relevant signal processes contributing to the 2b+$\not{E}_T$ final state consist of the following production channels of the heavy CP-even Higgs, $H$ such as

$$e^-e^+ \to b\bar{b}H, \;\; t\bar{t}H, \; ZH, \; HA$$

The production cross sections for all the signal processes are summarized in Table 8. As we see from Table 8, $HA$ and $t\bar{t}H$ are dominant signal processes while contributions from $ZH$ and $b\bar{b}H$ are negligible.

We study the final state of $2b$+$\not{E}_T$ with the missing energy arising from $H \to \chi\bar{\chi}$ at an unpolarized $e^+e^-$ collider for an integrated luminosity of $\mathcal{L} = 5$ ab$^{-1}$. Dominant SM background contributions arise from

- $b\bar{b}$ (for misidentification of decay products of $b$-quark along with missing energy from $b$ decays)
- $b\bar{b}\nu\bar{\nu}$ (including both on-shell and off-shell contribution from $Z$ boson decay as well as $\nu\bar{\nu}h$ contribution)
- $Z(\to b\bar{b})Z(\to \nu\bar{\nu})$
- $hZ$ ($h \to b\bar{b}$, $Z \to \nu\bar{\nu}$)
- $t\bar{t}Z$, ($Z \to \nu\bar{\nu}$, $t \to bW^+$) for misidentified leptons/jets from $W$ bosons.
- Leptonic $t\bar{t}$ for misidentified leptons or semi-leptonic $t\bar{t}$ decays with missing energy arising from the $b$ decaying leptonically.
- $WWZ$
- $ZZZ$

For the backgrounds processes, e.g., $t\bar{t}$, $t\bar{t}Z$ we generate the leptonic mode. For the other SM backgrounds, the $Z(\to bb\,)Z(\to \nu\bar{\nu})$, $h(\to bb\,)Z(\to \nu\bar{\nu})$, $W(l\nu)W(l\nu)Z(\to b\bar{b})$ modes are generated. The signal and background production cross section in the missing energy final state are summarized in Table 8 and 9.

#### *5.5.8 Pre-selection cuts*

The cuts used for identifying the reconstructed objects after reconstruction with the default ILD card in `Delphes` based on Ref. [84] are summarized below and denoted as **C0**,

- Leptons are reconstructed with a minimum transverse momentum $p_T > 10$ GeV and pseudorapidity $|\eta| < 2.5$.
- Photons are reconstructed with $p_T > 10$ GeV and $|\eta| < 2.5$.
- All (b)-jets are identified with minimum $p_T > 20$ GeV with pseudorapidity $|\eta| < 3.0$.

#### *5.5.9 Signal region C: $2b + \not{E}_T$*

We investigate the prospects of the final state consisting of two b-jets along with missing energy and perform a signal-to-background analyses using the cuts **C1-C6** as follows (Fig. 7),

- **C1**: The final state consists of two b-jets and no leptons or photons.
- **C2**: The leading b-jet has transverse momentum $p_T > 100$ GeV and sub-leading b-jet has $p_T > 80$ GeV. The hard $p_T$ cuts on the b-jets help to reduce backgrounds from SM backgrounds from $Z$ and $h$ bosons.
- **C3**: The invariant mass of the two b-jets within the mass window 80 GeV $< M_{b_1 b_2} <$130 GeV is rejected to remove contributions from $Z$ and $h$ bosons.

**Table 8** Unpolarized production cross sections at leading order (LO) for **BP1** at a linear $e^+e^-$ collider for $\sqrt{s} = 3$ TeV

| Benchmark | Process | cross sections (in fb) at $\sqrt{s} = 3$ TeV |
|---|---|---|
| **BP1** | $HA$ | 0.718 |
| | $t\bar{t}H$ | 0.403 |
| | $b\bar{b}H$ | $5.32\times10^{-7}$ |
| | $ZH$ | $6.5\times10^{-7}$ |

**Table 9** Unpolarized background cross sections for SM backgrounds with leptonic decay models for $t$ quarks, $W$ boson, invisible decay of $Z$, $Z \to b\bar{b}$ and $h \to b\bar{b}$

| SM Process | cross sections (in fb) |
|---|---|
| $ZZ$ | 1.54 |
| $WW$ | 22.12 |
| $b\bar{b}\nu\bar{\nu}$ | 40.84 |
| $b\bar{b}$ | 10.17 |
| $t\bar{t}$ (leptonic) | 0.87 |
| $t\bar{t}$(semi-leptonic) | 5.2 |
| $t\bar{t}Z$ | 0.0149 |
| $hZ$ | 0.23 |
| $WWZ$ | 0.23 |
| $ZZZ$ | 0.0151 |

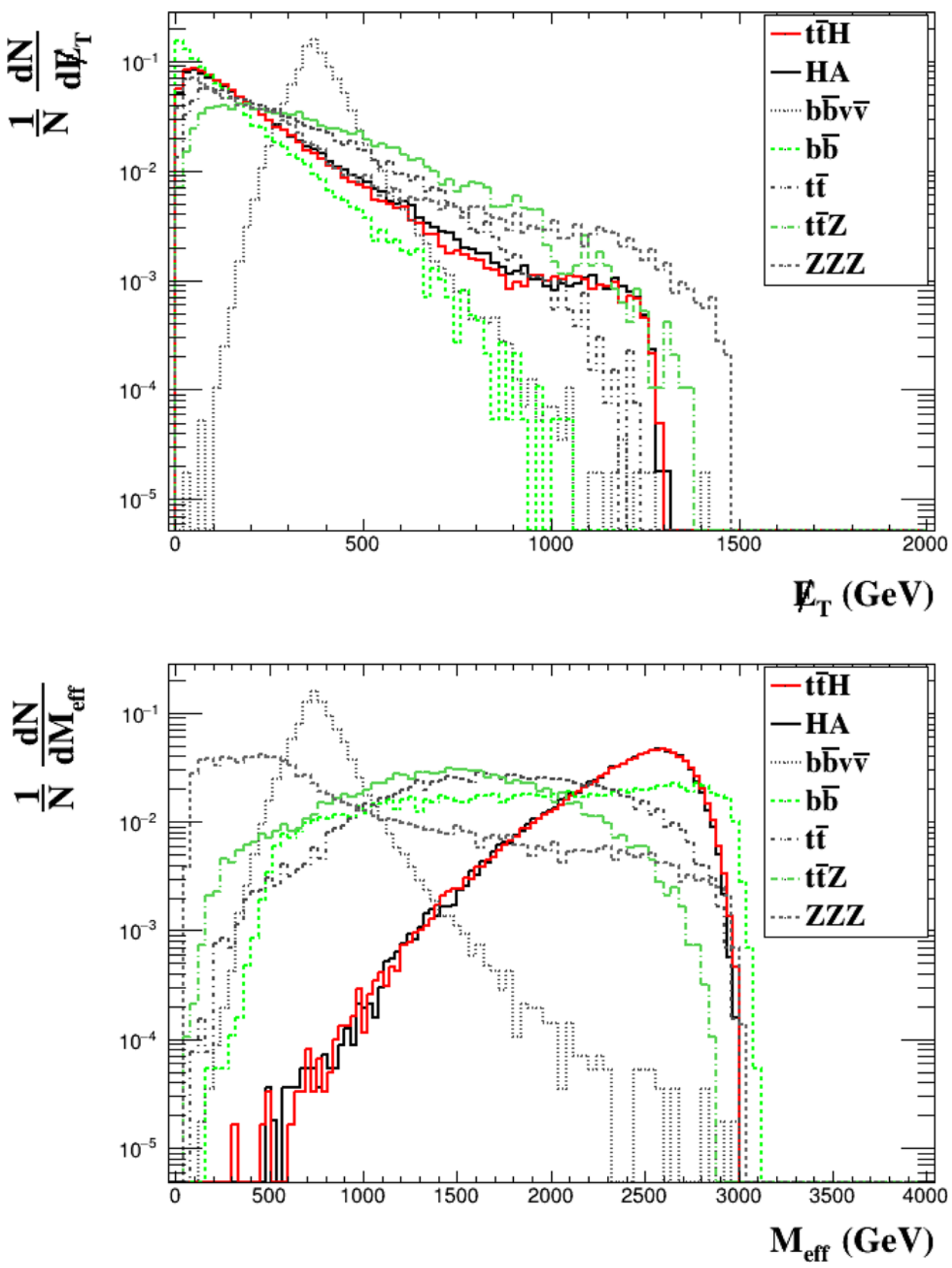

**Fig. 7** Normalized distribution of the missing transverse momentum ($\not{E}_T$) (top) and effective mass($M_{eff}$) (bottom) for signal vs. some backgrounds after cut **C1**. Concerning the $t\bar{t}$ background the leptonic $t\bar{t}$ contribution has been chosen

- **C4**: Since the dark matter is heavy, we demand a large cut on the effective mass $M_{eff} > 1.2$ TeV where $M_{eff} = \sum_i (p_{T_i}) + \not{E}_T$.
- **C5**: Further, we demand $\not{E}_T > 650$ GeV on the final state which reduces the dominant SM backgrounds.
- **C6**: The $\Delta\Phi$ between two b-jets is significantly different for the signal and background from $b\bar{b}$ where the b-jets are mostly back-to-back. We demand $\Delta\Phi(b_1, b_2) < 1.60$. This also reduces the backgrounds from $b\bar{b}$ as well as from $t\bar{t}$ and $t\bar{t}Z$ sharply.

The number of signal and background events after applying the cuts **C1-C6** (at $\mathcal{L} = 5$ ab$^{-1}$) is summarized in Table 10. We observe that the large $M_{eff}$ and $\not{E}_T$ cut are instrumental in reducing SM backgrounds sharply as shown in Fig. 7. Furthermore, the variable $\Delta\Phi(b_1, b_2)$ also reduces contributions from $b\bar{b}$ owing to the fact that the jets in the background are back to back while in the signal they are more collimated as shown in Fig. 9. It also reduces the backgrounds from $t\bar{t}$ and $t\bar{t}Z$ sharply which are also peaked toward higher values of $\Delta\Phi$ as compared to the signal. We also plot the invariant mass of the two b-jets in Fig. 8 and observe that the SM backgrounds from $Z$ and $h$ are peaked at the resonance masses but for the signal the peak is broader since the parent particle is heavy. However, the irreducible background $b\bar{b}\nu\bar{\nu}$ has a similar shape compared to the signal.

**Fig. 8** Normalized distribution of the invariant mass of the two b-jets $M_{b\bar{b}}$ for signal vs. background after cut **C1**. Concerning the $t\bar{t}$ background the leptonic $t\bar{t}$ contribution has been chosen

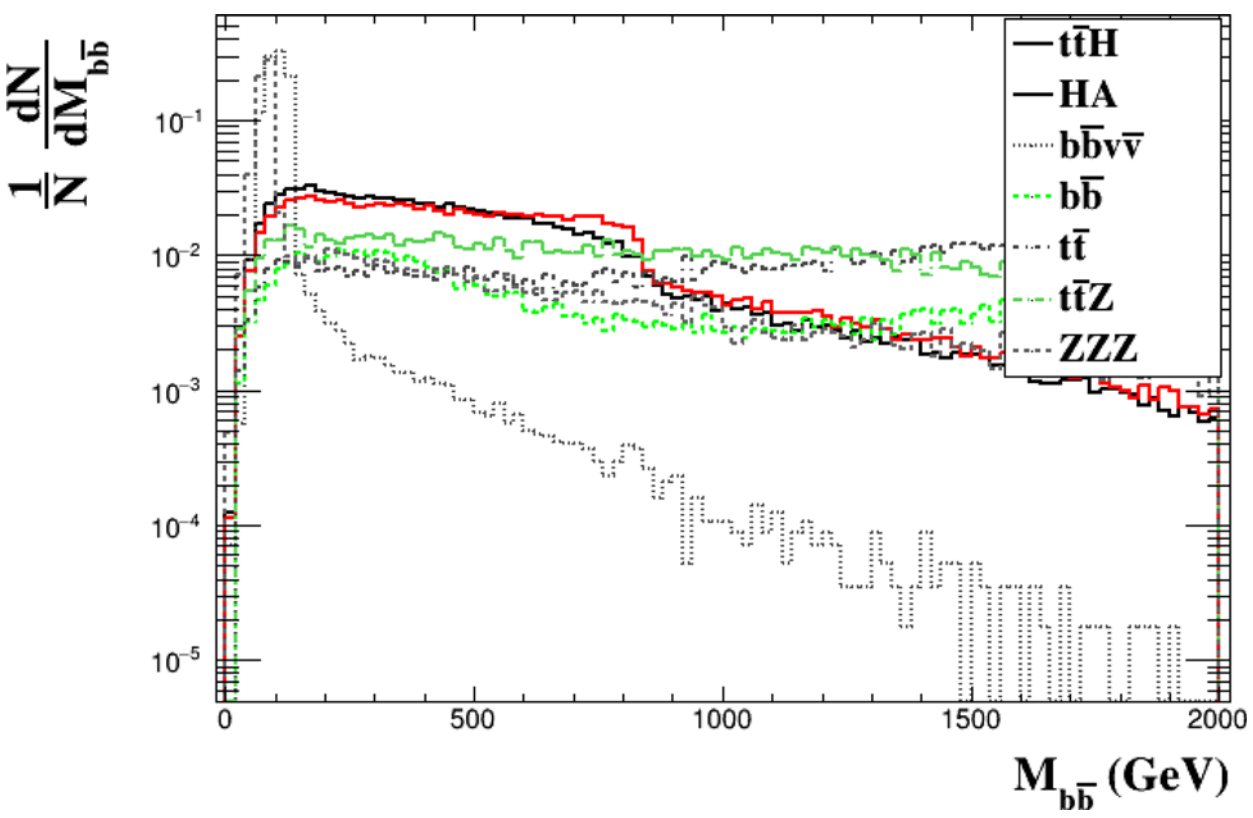

**Fig. 9** Normalized distribution of $\Delta\Phi$ separation between the two final state b-jets after cut **C1**. Concerning the $t\bar{t}$ background the leptonic $t\bar{t}$ contribution has been chosen

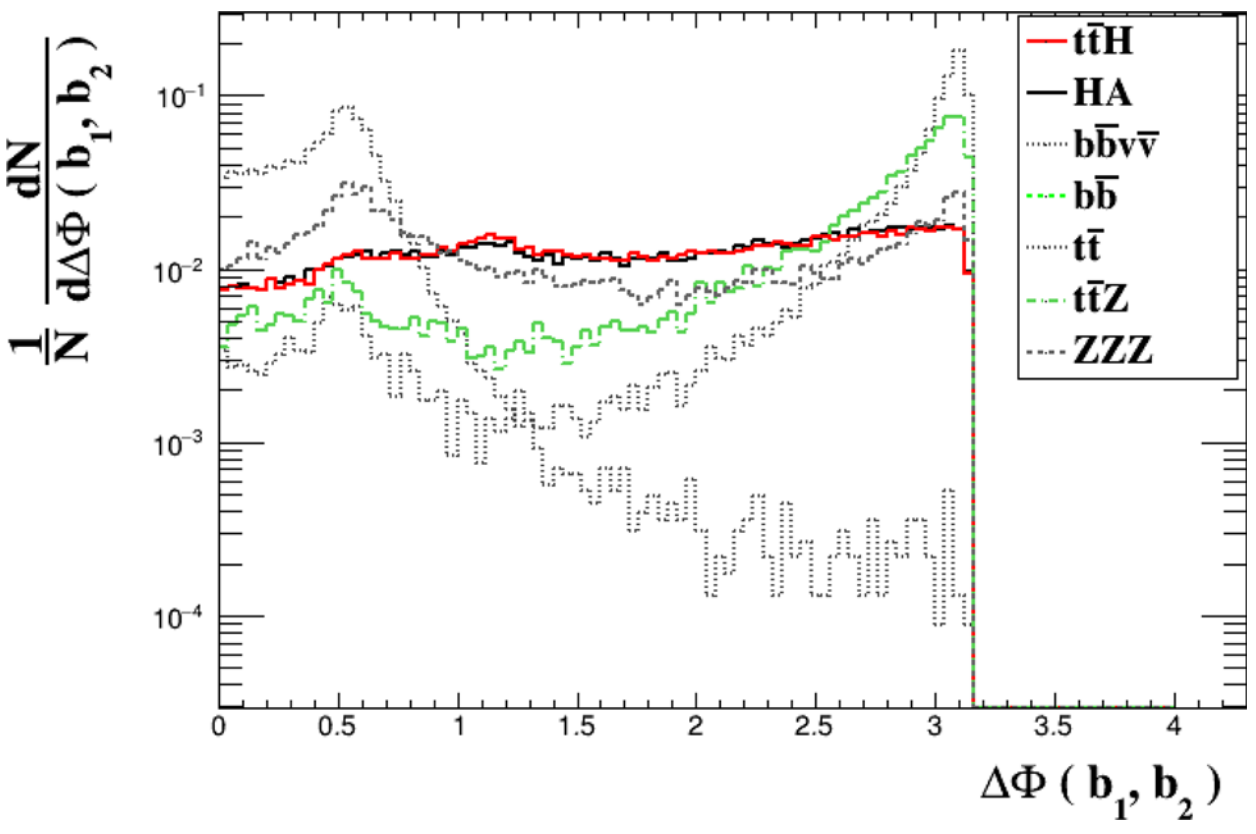

Therefore, excluding the mass window 80 GeV $< M_{b_1 b_2} <$130 GeV, removes the resonant backgrounds from $Z$ and $h$ bosons, however with retaining backgrounds involving b-jets associated with neutrinos and also from $t\bar{t}$ associated backgrounds (Fig. 9).

We discuss the signal significance only for unpolarized electron and positron beams. The statistical significance ($\mathcal{S}$) of the signal ($s$) over the total SM background ($b$) is calculated using [85, 86]

$$\mathcal{S} = \sqrt{2 \times \left[(s+b)\ln(1+\frac{s}{b}) - s\right]} \tag{39}$$

where $s$ and $b$ are the total signal and background event numbers after the cuts **C1-C6**. We use this expression for the statistical significance since the background events are not overwhelmingly large compared to the signal and the limit $b >> s$ is not fully accurate.[3] We observe that for **BP**, owing to the small invisible decay branching ratio of the heavy Higgs, i.e., $H \to \chi\bar{\chi} \sim 1.94\%$, one can obtain 1.36 $\sigma$ signal at the integrated luminosity of 5 ab$^{-1}$.

For a higher branching fraction, as for benchmark **BP2** with an increased invisible branching, $BR(H \to \chi\chi) = 4.67\%$, the signal significance improves to ∼2.31 $\sigma$ at leading order (LO). Note that we have used unpolarized incoming beams for this study. It is well known that the use of a right-handed electron and left-handed positron beam polarization can significantly suppress SM backgrounds [79] and can be effectively used to improve the signal significance as proposed at the upcoming ILC with the possibility of achieving beam polarizations (80%, 30-60%)) for $(e^-, e^+)$. With the CLIC experiment potentially targeting higher center-of-mass energies but with electron polarization only, one may hope to further enlarge the discovery range.

## 6 The U(1) breaking case

We work in the two-component framework where the complex scalar S is decomposed into its components $h_s$ and $a_s$, both of which are odd under $Z_2'$. In this case, both $h_s$ and $a_s$ contribute to the relic density. From sec. 2, recall eq. 23 and 24 for the squared-masses of $h_s$ and $a_s$ are

$$m_{h_s}^2 = \frac{1}{2}(m_S^2 + m_S^{2\prime} + (\lambda_1' + 2\lambda_4')v_1^2 + (\lambda_2' + 2\lambda_5')v_2^2),$$
$$m_{a_s}^2 = \frac{1}{2}(m_S^2 - m_S^{2\prime} + (\lambda_1' - 2\lambda_4')v_1^2 + (\lambda_2' - 2\lambda_5')v_2^2)$$

and eq. 25

$$\Delta^2 = |m_{h_s}^2 - m_{a_s}^2| = m_S^{2\prime} + 2\lambda_4' v_1^2 + 2\lambda_5' v_2^2$$

where $\Delta^2$ is the squared mass difference of the two dark states. The mixing between $h_s$ and $a_s$ is absent in this case owing to the form of the scalar potential by construction since the mixed $h_s a_s$ terms cancel when the Hermitian conjugate terms are included as is done to ensure the potential is Hermitian. Therefore this leads to a multi-component dark sector with two dark sector particles which are not identical in mass leading to a two component DM scenario. The couplings of $h_s$ and $a_s$ with the Higgs bosons are presented in Eqs. 19-22. The portal couplings $\lambda_1'$, $\lambda_2'$, $\lambda_4'$ and $\lambda_5'$ are instrumental in determining the coupling of the Higgs bosons to the dark sector particles and affect the DM phenomenology crucially.

[3] In the limit $b >> s$, Eq. 39 reduces to $S = \frac{s}{\sqrt{b}} \simeq 1.41\,\sigma$ ( for $\mathcal{S} = \frac{s}{\sqrt{s+b}} \simeq 1.34\,\sigma$).

**Table 10** The cut-flow table showing the change in the number of events for the benchmark **BP1** for unpolarized electron and positron beams at $\sqrt{s} = 3$ TeV at $\mathcal{L} = 5$ ab$^{-1}$. The cuts (**C0-C6**) are defined in the text in Sect. 5.2. The '-' denotes results with less than 1 event at an integrated luminosity $\mathcal{L} = 5$ ab$^{-1}$

| Process | C0 | C1 | C2 | C3 | C4 | C5 | C6 |
|---|---|---|---|---|---|---|---|
| $HA$ | 3588 | 729.9 | 494.4 | 488.2 | 486.7 | 17.42 | 11.6 |
| $t\bar{t}H$ | 2005 | 375.7 | 233.4 | 230.2 | 229.2 | 9.20 | 5.5 |
| **BP1** | | | | | | | 17.1 |
| $b\bar{b}\nu\bar{\nu}$ | 204200 | 35159 | 15738 | 2040.9 | 330.3 | 147.6 | 124.3 |
| $b\bar{b}$ | 50850 | 9514.5 | 8432.5 | 8387.2 | 6697.5 | 65.6 | 4.07 |
| $ZZZ$ | 75 | 10.89 | 3.75 | 3.07 | 1.5 | 0.51 | 0.28 |
| $WWZ$ | 1163 | 42.8 | 3.14 | 1.1 | 0.14 | 0.02 | – |
| $t\bar{t}Z$ | 74 | 7.1 | 5.68 | 5.6 | 4.04 | 0.71 | 0.35 |
| $t\bar{t}$ (semi-leptonic) | 25955 | 3846.1 | 2843.9 | 2818.8 | 2500.6 | 338.5 | 16.61 |
| $t\bar{t}$ (leptonic) | 4328 | 565.4 | 481.5 | 478.3 | 401.9 | 29.65 | 1.13 |
| $WW$ | 110620 | 0.55 | 0.28 | 0.28 | - | - | – |
| $hZ$ | 1131 | 6.8 | 1.26 | 0.023 | - | - | – |
| $ZZ$ | 7699 | 167.9 | 42.81 | 13.0 | - | - | – |
| Total background | | | | | | | 146.4 |
| Significance | | | | | | | 1.36 |

For simplicity, we redefine the singlet scalar potential as

$$\begin{aligned} V_S^m = &\frac{m_{DM_1}^2}{4} h_s^2 + \lambda_1^{m\prime\prime} h_s^4 \\ &+ \lambda_1^{m\prime} h_s^2 \Phi_1^\dagger \Phi_1 + \lambda_2^{m\prime} h_s^2 \Phi_2^\dagger \Phi_2 \\ &+ \frac{m_{DM_2}^2}{4} a_s^2 + \lambda_3^{m\prime\prime} a_s^4 \\ &+ \lambda_4^{m\prime} a_s^2 \Phi_1^\dagger \Phi_1 + \lambda_5^{m\prime} h_s^2 \Phi_2^\dagger \Phi_2 \\ &+ \lambda_2^{m\prime\prime} h_s^2 a_s^2, \end{aligned} \tag{40}$$

where the superscript $m$ stands for multi-component and

$$\begin{aligned} m_{DM_1}^2 &= 2(m_S^2 + m_S^{2\prime}), \\ m_{DM_2}^2 &= 2(m_S^2 - m_S^{2\prime}), \\ \lambda_1^{m\prime} &= \frac{\lambda_1' + 2\lambda_4'}{2}, \\ \lambda_2^{m\prime} &= \frac{\lambda_2' + 2\lambda_5'}{2}, \\ \lambda_4^{m\prime} &= -\frac{\lambda_1' - 2\lambda_4'}{2}, \\ \lambda_5^{m\prime} &= -\frac{\lambda_2' - 2\lambda_5'}{2}, \end{aligned}$$

**Table 11** List of parameters varied. The other parameters, $\lambda_1 = 0.21$, $\lambda_2 = 0.26$, $\lambda_3 = 0.199$, $\lambda_4 = -0.1$, $\lambda_5 = 0.10$, $m_{12}^2 = 10^5$, $\tan\beta = 6.8$, $\lambda_3^{m''} = 0.0167$, $\lambda_1^{m'} = 2.0$, $\lambda_4^{m'} = 0.2$, $\lambda_5^{m'} = 0.01$ and $\lambda_2^{m'} = 0.01$, $\lambda_1^{m''} = 0.0$ and $\lambda_2^{m''} = 0.083$ are kept fixed throughout the scan

| Parameters | Range (GeV$^2$) |
|---|---|
| $m_{DM_1}^2$ | 0-1000000 |
| $m_{DM_2}^2$ | 0-1000000 |

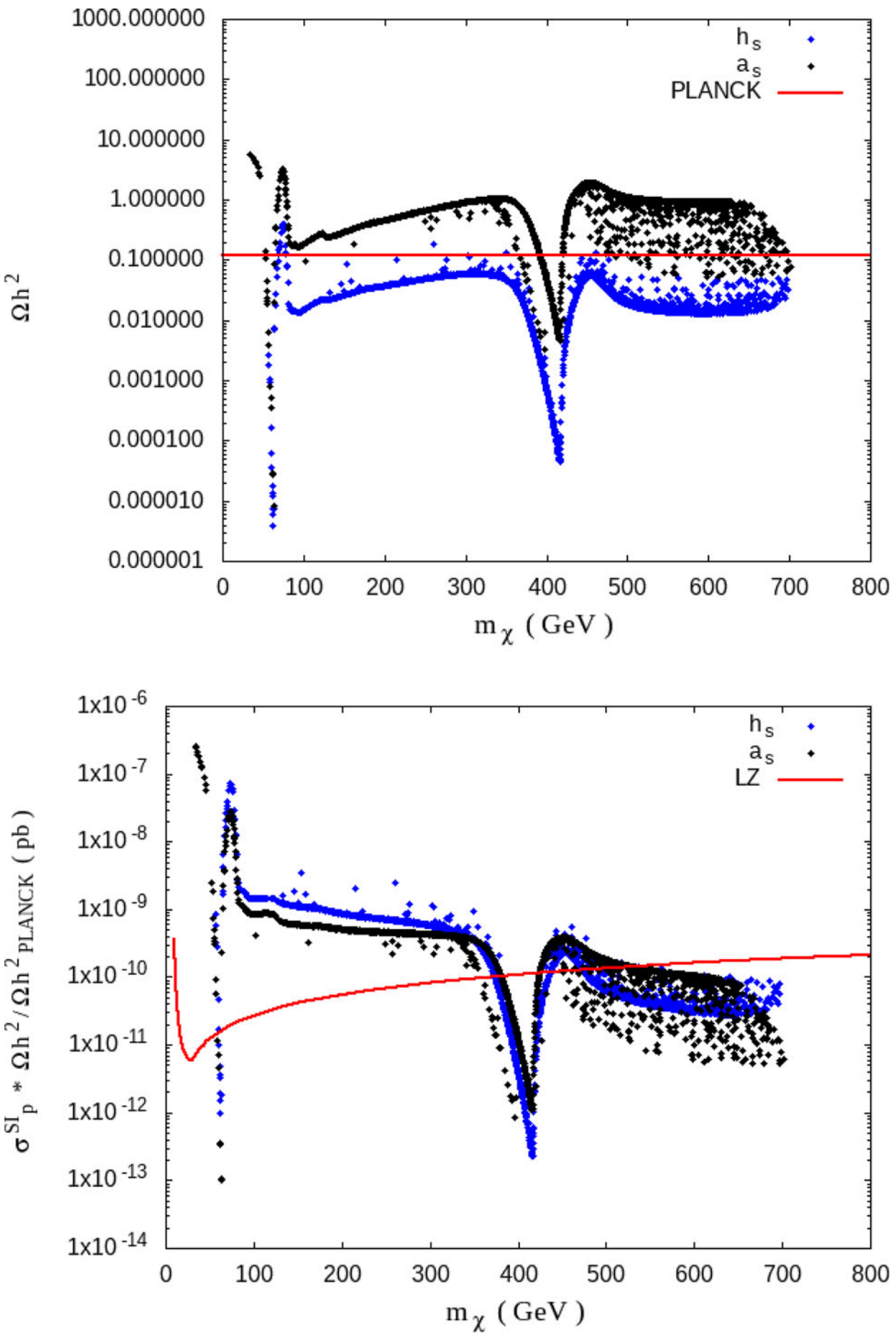

**Fig. 10** Variation of the total relic density (top) and spin independent direct detection cross section (bottom) for cases when $h_s = \chi$ (blue) and $a_s = \chi$ (black) vs. the mass of the lightest DM candidate $m_\chi$ where $\chi$ is the lightest dark sector particle

$$\lambda_1^{m''} = \frac{\lambda_3''}{16} + \frac{\lambda_1''}{48} + \frac{\lambda_2''}{12},$$
$$\lambda_2^{m''} = \frac{\lambda_3''}{8} - \frac{\lambda_1''}{48},$$
$$\lambda_3^{m''} = \frac{\lambda_3''}{16} + \frac{\lambda_1''}{48} - \frac{\lambda_2''}{12}. \quad (41)$$

We fix the parameters as in Table 11 and perform a random scan for $m_{DM_1}$ and $m_{DM_2}$.

Figure 10 shows the variation of the total relic density versus the mass of the lightest dark sector particle denoted as $m_\chi$. The blue points represent the cases where $h_s$ is the lightest DM candidate, while the black points represent the case where $a_s$ is the lightest DM candidate.

We observe that in both cases, there are allowed regions from the relic density constraint from PLANCK for $m_\chi$ near the 125 GeV Higgs and the heavy CP-even Higgs resonances as shown by the dips near $m_\chi \sim \frac{m_h}{2}$ and $\frac{m_H}{2}$, respectively. For cases with $h_s$ as the lightest DM candidate, the parameter space remains relatively less constrained with $m_\chi > 100$ GeV allowed from the relic density upper limit while for the case with $a_s$ as the lightest DM candidate, only parameter points near the Higgs resonances and beyond $m_\chi > 450$ GeV are allowed by PLANCK constraint. This follows from the fact that $h_s$ couples more strongly to the DM pair compared to $a_s$ for the parameters chosen for this scan. This is due to the fact that $\lambda_1^{m\prime} > \lambda_4^{m\prime}$ and $\lambda_2^{m\prime} > \lambda_5^{m\prime}$ as seen from eq. 41 for the choice of parameters in this scan. When $h_s$ is the lightest DM candidate, it dominates the contribution to the relic density while the heavier DM candidate suffers due to Boltzmann suppression. When $a_s$ is the lightest DM candidate, since it couples weakly to the Higgses, the total relic density is higher compared to when $h_s$ is the lightest DM candidate. The constraint from spin-independent direct detection search from LZ, stringently constrains both the cases with $h_s$ and $a_s$ allowing points near the Higgs resonances and for heavy dark matter where $m_\chi \geq 475$ GeV.

Thus, in summary, we observe that in this case, the parameter space is stringently restrained from relic density except for cases when $h_s$ is the lightest DM candidate near the Higgs resonances and for $m_\chi > 100$ GeV. For cases with $a_s$ as the lightest DM particle, the relic density constraint allows the Higgs resonance regions and heavy dark matter for $m_\chi > 450$ GeV due to the weaker couplings of $a_s$ to the Higgses compared to that of $h_s$. Direct detection constraints, however, stringently constrain both scenarios and favor parameter spaces near the resonances and heavy dark matter masses beyond the heavy Higgs resonance region where $m_\chi \geq 475$ GeV.

### 6.1 Collider analyses

We choose a sample benchmark with the Higgs sector similar to **BP1** and present an estimate for observing a sample benchmark **BP3** (as shown in Table 12) at an $e^+e^-$ collider. For BP3, the pseudoscalar $a_s$ is the lightest dark sector particle and contributes to the DM relic density while the heavier dark sector particle $h_s$ does not. We obtain a simplistic estimate of the signal significance by rescaling the results from the collider analyses for the signal region C for **BP1** using the new branching fraction of **BP3** to the lightest dark matter candidate to estimate the signal significance. Using the cut efficiency of signal region C analysis, we obtain $\simeq$ 34 events for **BP3** leading to a signal significance of $\sim 2.74\sigma$ at an $e^+e^-$ collider at $\sqrt{s} = 3$ TeV at 5ab$^{-1}$. We leave a more detailed collider study subject to all theoretical and experimental constraints in a forthcoming study.

## 7 Summary and conclusions

We consider the Type II two-Higgs-doublet model extended with a complex singlet scalar (2HDMS), motivated to address several open issues in nature. The softly broken $Z_2$-symmetric 2HDM scalar potential is augmented with a complex scalar symmetric under $Z_2'$, stabilizing the dark matter candidate. Since the complex scalar singlet does not develop a vacuum expectation value, the Higgs spectrum is the same as in the 2HDM. In the CP-conserving scenario, the CP-even Higgs bosons act as a portal to the dark matter. We discuss both the U(1) symmetric and U(1) breaking cases for our model. In the U(1) symmetric case, the complex singlet may be treated as a dark matter candidate. We explore the parameter space allowed by current experimental data from dark matter, flavour physics and collider constraints from the SM-like Higgs as well as searches for the heavy Higgs bosons. We observe that direct detection results stringently constrain the parameter space and require low values of $\lambda_2'$. Furthermore, we compare our results with the real singlet extended 2HDM and observe that the contributions to the relic density are much larger for the complex scalar DM over the real scalar DM while the direct detection cross section are very similar in both extensions.

The presence of the singlet also leads to new decay modes for the Higgs bosons namely into a pair of dark matter particles. Such a final state will lead to the signature of missing energy at colliders. We choose a representative benchmark **BP1** with $m_H \simeq 830$ GeV consistent with all experimental data in order to demonstrate the prospects of observing such a signal at HL-LHC and at future $e^+e^-$ colliders. In this case, the constraints from direct detection data as well as the competing fermionic decays of the heavy Higgs bosons stringently constrain the invisible decay $H \to \chi\bar{\chi} \sim 1.94\%$. Due to the small branching fraction and the heavy mass of $H$, rather low cross sections HL-LHC are obtained for gluon fusion and vector boson fusion process. However, one can achieve $\sim 6\sigma$ significance for **BP1** at HL-LHC. For VBF process, the signal is significantly suppressed and beyond the reach of HL-LHC due to the low production cross section of the heavy Higgs. However, one has a better control on the background at an e+e- collider due to the precise initial energy and the clean environment as compared to the HL-LHC. We perform a signal-to-background analyses at the $e^+e^-$ collider with $\sqrt{s} = 3$ TeV with unpolarized beams and derive that the $2b + \not{E}_T$ channel is observable with a signal significance of 1.36 $\sigma$ at an integrated luminosity $\mathcal{L} = 5$ ab$^{-1}$. For a higher branching fraction of the heavy Higgs to dark matter, the significance improves to $\sim 3.65\ \sigma$. One should note, however, that still beam polarization has not been applied for the current study, which may lead to higher significance for **BP1** and **BP2**.

We also study the dark matter phenomenology for the U(1) breaking case and discuss the constraints from the relic and direct detection data and choose a benchmark example **BP3** to demonstrate the scenario. We observe that one can reach a higher invisible branching in this case $\simeq 3.9\%$ and signal significance $\sim 2.81\sigma$ is within the reach at a future e$^+$ e$^-$ factory. However, we have not yet included beam polarization effects yet. We leave a further optimized study for the linear collider for a future study.

**Table 12** Relevant parameters of the benchmark used for the study

| Parameters | **BP3** |
|---|---|
| $\lambda_1$ | 0.21 |
| $\lambda_2$ | 0.26 |
| $\lambda_3$ | 0.25 |
| $\lambda_4$ | –0.1 |
| $\lambda_5$ | 0.1 |
| $\lambda_1^{\prime\prime m}$ | 0.01 |
| $\lambda_2^{\prime\prime m}$ | 0.0001 |
| $\lambda_3^{\prime\prime m}$ | 0.001 |
| $\lambda_1^{\prime m}$ | 2.0 |
| $\lambda_2^{\prime m}$ | 0.1 |
| $\lambda_4^{\prime m}$ | –0.1 |
| $\lambda_5^{\prime m}$ | -0.1 |
| $m_{12}^2(\mathrm{GeV}^2)$ | $1.0\times10^5$ |
| $m_{DM_1}^2(\mathrm{GeV}^2)$ | 300000 |
| $m_{DM_2}^2(\mathrm{GeV}^2)$ | 380000 |
| $\tan\beta$ | 6.8 |
| $\cos(\alpha-\beta)$ | –0.00015 |
| $m_h$ (GeV) | 125.6 |
| $m_H$(GeV) | 833.5 |
| $m_A$(GeV) | 829.8 |
| $m_{H^\pm}$ (GeV) | 833.5 |
| $m_{a_s}$ (GeV) | 398.12 |
| $m_{h_s}$ (GeV) | 428.88 |
| $\mathrm{BR}(H\to a_s a_s)$ | $3.9\times10^{-2}$ |
| $\Omega h^2$ | $1.03\times10^{-3}$ |
| $\sigma_p^{SI} * \frac{\Omega h^2}{\Omega h^2_{PLANCK}}$ $(\times10^{11}$ pb) | 2.66 |
| $\sigma_n^{SI} * \frac{\Omega h^2}{\Omega h^2_{PLANCK}}$ $(\times10^{11}$ pb) | 2.79 |

**Acknowledgements** JD and GMP acknowledge support by the Deutsche Forschungsgemeinschaft (DFG, German Research Foundation) under Germany's Excellence Strategy EXC 2121 "Quantum Universe"- 390833306. The authors thank G.Belanger, A.Pukhov from the `micrOMEGAs` team, J.Heisig, O.Mattalear and C.Arina from the `maDDM` team for helpful correspondence. The authors also thank H.Bahl, S.Heinemeyer, C.Li, S. Paasch, T.Stefaniak and G.Weiglein for helpful discussions and help with `HiggsBounds` and `HiggsSignals` and J.Lahiri for useful discussions, comments and for help with using `EVADE`.

**Funding** Open Access funding enabled and organized by Projekt DEAL.

**Data availability** Data sets generated during the current study are available from the corresponding author on reasonable request.

**Table 13** The couplings of the quarks in Type II 2HDM[30]

| Higgs bosons | $C_u$ | $C_d$ | $C_s$ |
|---|---|---|---|
| $h$ | $\cos\alpha/\sin\beta$ | $-\sin\alpha/\cos\beta$ | $-\sin\alpha/\cos\beta$ |
| $H$ | $\sin\alpha/\sin\beta$ | $\cos\alpha/\cos\beta$ | $\cos\alpha/\cos\beta$ |

## Appendix

A. Analytical computation of direct detection cross section

We compute the direct detection cross section. The DM-quark amplitude of the tree-level Feynman diagram shown in Fig. 4 for zero momentum transfer is

$$\mathcal{M} = \sum_{h_i} \lambda_{h_i SS^*} \frac{1}{-m_{h_i}^2} \lambda_{h_i q\bar{q}}, \tag{42}$$

where $h_i = h,\ H$. Recall,

$$\lambda_{hSS^*} = \frac{v}{\sqrt{1+\tan^2\beta}}(\lambda_1' \sin\alpha - \lambda_2' \tan\beta\cos\alpha),$$
$$\lambda_{HSS^*} = -\frac{v}{\sqrt{1+\tan^2\beta}}(\lambda_1' \cos\alpha - \lambda_2' \tan\beta\sin\alpha)$$

and

$$\lambda_{h_i q\bar{q}} = \frac{m_q}{v} C^{h_i}$$

is the Yukawa coupling of the quark to the Higgs bosons. For Type II 2HDM[30], the Higgs couplings to the fermions are summarized in Table 13.

In order to compute the DM-nucleon amplitude, the nucleon form factors $f_N$ need to be folded into the quark-DM amplitude $\mathcal{M}$. The form factors $f^N$ for $N = p,\ n$ are[14]

$$f_N = \frac{m_N}{2m_\chi}\Big(\sum_{q=u,d,s} f_{Tq}^N \frac{\mathcal{M}}{m_q} + \frac{2}{27} f_{TG}^N \sum_{q=c,b,t} \frac{\mathcal{M}}{m_q}\Big), \tag{43}$$

where the first term is due to the contribution of the light valence quarks and the second term due to the gluonic form factor $f_{TG}^N$ defined as,

$$f_{TG}^N = 1 - \sum_{q=u,d,s} f_{Tq}^N. \tag{44}$$

Folding in the form factors above and the phase space of the DM-nucleon scattering, the DM-nucleon cross sections, for the proton and neutron, respectively, are[14, 87],

$$\sigma_{SI}^p = \frac{4\mu_p^2}{\pi}(Zf_p)^2 \tag{45}$$

and

$$\sigma_{SI}^n = \frac{4\mu_n^2}{\pi}((A-Z)f_n)^2 \tag{46}$$

where $\mu_{p,n} = \frac{m_\chi m_{p,n}}{(m_\chi + m_{p,n})}$.